\documentclass[aps,prb,twocolumn,longbibliography]{revtex4-1}

\usepackage{amsmath}
\usepackage{amsfonts}
\usepackage{mathtools}
\usepackage{braket}
\usepackage{bm}
\usepackage{array}
\usepackage{graphicx}
\usepackage[dvipsnames]{xcolor}

\usepackage{hyperref}
\usepackage{bbold}

\hypersetup{
colorlinks=true,
linkcolor=blue,
citecolor=blue,
filecolor=magenta,
urlcolor=cyan
}

\definecolor{oddprimary}{HTML}{e6194b}%Red
\definecolor{phsdesc}{HTML}{1333a8}%Blue
\definecolor{trsdesc}{HTML}{1333a8}%Blue
\definecolor{odddescphs}{HTML}{f58231}%Purple
\definecolor{odddesctrs}{HTML}{f58231}%Orange
\definecolor{even}{HTML}{1c942b}%Green

\DeclareMathOperator{\Tr}{Tr}
\DeclareMathOperator{\Dim}{Dim}
\DeclareMathOperator{\sgn}{sgn}

\newcommand{\papertitle}{Classification of topological insulators and superconductors out of equilibrium}
\newcommand{\tcm}{T.C.M. Group, Cavendish Laboratory, University of Cambridge, JJ Thomson Avenue, Cambridge, CB3 0HE, U.K.}

\DeclareSymbolFont{sfletters}{OML}{cmbrm}{m}{it}

\newcommand{\matr}[1]{\mathsf{#1}}
\DeclareMathSymbol{\matrrho}{\mathord}{sfletters}{"1A}
\newcommand{\diff}{\mathrm{d}}

\newcommand{\noiseavg}[1]{\braket{#1}_\text{noise}}

\DeclareFontFamily{OMX}{MnSymbolE}{}
\DeclareFontShape{OMX}{MnSymbolE}{m}{n}{
	<-6>  MnSymbolE5
	<6-7>  MnSymbolE6
	<7-8>  MnSymbolE7
	<8-9>  MnSymbolE8
	<9-10> MnSymbolE9
	<10-12> MnSymbolE10
	<12->   MnSymbolE12}{}
\DeclareSymbolFont{mnlargesymbols}{OMX}{MnSymbolE}{m}{n}
\SetSymbolFont{mnlargesymbols}{bold}{OMX}{MnSymbolE}{b}{n}
\DeclareMathDelimiter{\llangle}{\mathopen}{mnlargesymbols}{'164}{mnlargesymbols}{'164}
\DeclareMathDelimiter{\rrangle}{\mathclose}{mnlargesymbols}{'171}{mnlargesymbols}{'171}

\begin{document}

\title{\papertitle}
\author{Max McGinley}
\affiliation{\tcm}
\author{Nigel R. Cooper}
\affiliation{\tcm}

\date{\today}

\begin{abstract}
	We establish the existence of a topological classification of many-particle quantum systems undergoing unitary time evolution. The classification naturally inherits phenomenology familiar from equilibrium -- it is robust against disorder and interactions, and exhibits a non-equilibrium bulk-boundary correspondence, which connects bulk topological properties to the entanglement spectrum. We explicitly construct a non-equilibrium classification of non-interacting fermionic systems with non-spatial symmetries in all spatial dimensions (the `ten-fold way'), which differs from its equilibrium counterpart. Direct physical consequences of our classification are discussed, including important ramifications for the use of topological zero-energy bound states in quantum information technologies.
	%We construct a topological classification of non-interacting fermionic many-body wavefunctions far from equilibrium. The symmetry properties of states undergoing unitary time evolution can differ from the symmetries of the initial state and the Hamiltonian, which leads to a reduction of the equilibrium classification for topological states that rely on certain symmetries. We systematically determine the group of topologically distinct non-equilibrium wavefunctions as a function of spatial dimension and the symmetry class of the Hamiltonians. Our non-equilibrium classification can be understood both through bulk properties and through the study of certain boundary theories, as evidenced in the entanglement spectrum.
\end{abstract}

\maketitle

\section{Introduction}

Topology has become one of the most prevalent concepts in condensed matter physics. In mathematics, the objects of interest for topologists are structures that remain invariant under continuous deformations of the underlying system. Likewise, \textit{physical} properties that do not change under continuous deformation of the underlying Hamiltonian are of interest in physics. In condensed matter, these topological properties are naturally robust to system imperfections and in some cases lead to accurately quantised responses, exemplified by the transverse conductance in the integer quantum Hall effect (IQHE) \cite{vonKlitzing1980,Thouless1982}.

Classifying different phases of matter according to their topological properties is one of the key aims of this field. Roughly speaking, gapped systems belong to different topological phases when their Hamiltonians cannot be continuously connected without meeting some topological quantum phase transition \cite{Kitaev2009}: a phase transition at zero temperature that is not characterized by spontaneous symmetry breaking \cite{Kosterlitz1973}. These topological phases are typically understood to be equilibrium (zero-temperature) properties of the Hamiltonian, which itself may be restricted by the symmetries of the system \cite{Xie2010}.

In this paper, rather than characterizing ground states of Hamiltonians, we ask whether there is a topological classification of non-equilibrium states $\ket{\Psi(t)}$ resulting from time-evolution under Hamiltonians with certain symmetries. We answer this question in the affirmative by developing a formalism for classifying wavefunctions which naturally generalizes the more familiar equilibrium classification. When applied to many-body systems undergoing unitary time evolution, this non-equilibrium topological classification emerges, which generally differs from that in equilibrium. The classification is relevant in physical settings where the Hamiltonian varies in time, e.g.~through a deliberate quench or in the presence of random noise.

We explicitly realise this non-equilibrium classification by building on the understanding of topology in equilibrium. Indeed many of the current theoretically \cite{Qi2011,Shiozaki2014,Senthil2015,Chiu2016,Kruthoff2017} and experimentally \cite{Stormer1999,Hasan2010,Ren2016} known topological phases of matter can be understood through unifying perspectives. One key feature common to many of these phases is the presence of symmetries: although two systems could in general be continuously connected, they may be disconnected in the subspace of states which respect certain symmetries -- such systems constitute the symmetry-protected topological (SPT) and symmetry-enriched topological (SET) phases \cite{Xie2010}. One important subset of SPT phases is the wide range of topological insulators and superconductors, which can be described by Hamiltonians that are quadratic in fermionic creation and annihilation operators (including mean-field Bogoliubov-de Gennes Hamiltonians), and have non-spatial symmetries. The equilibrium topology of these systems has been classified under the `ten-fold way' \cite{Qi2008,Schnyder2008,Kitaev2009,Ryu2010}.

Access to coherent dynamics in cold atom experiments \cite{Flaschner2018} has motivated recent studies of non-equilibrium physics in topological insulators and superconductors \cite{Foster2013,Sacramento2014,DAlessio2015,Caio2015,Wilson2016,Ulcakar2018}. A key aspect of these studies is the behaviour of topological bulk indices after a quench between Hamiltonians. In 2D Chern insulators (lattice analogues of the IQHE), it was shown that the relevant topological index (the Chern number) is preserved in time. However in Ref.~\onlinecite{McGinley2018} we demonstrated that in certain 1D systems, topological properties of symmetry-protected states can change out of equilibrium, even if the Hamiltonian never breaks the symmetries required to stabilise the phase in equilibrium.

The non-equilibrium classification which we construct in this paper unifies these previous results, and establishes a universal phenomenology which can be applied to any isolated quantum system. We explicitly derive the non-equilibrium analogue of the `ten-fold way', describing topological insulators and superconductors in arbitrary spatial dimension (Table \ref{tabClassification}). Specifically, given the spatial dimension and the set of symmetries possessed by the initial state and governing Hamiltonian, we provide an Abelian group, the elements of which represent states that remain topologically inequivalent after time-evolution. The classification naturally pertains to physical properties that are robust to disorder and interactions, and exhibits a bulk-boundary correspondence. As a first direct application of our results, we find that the classification can be used to predict instabilities of topological edge (Majorana) zero-modes to decoherence from an external fluctuating perturbation. We postulate that this connection between the preservation (destruction) of bulk topology in our non-equilibrium topological classification and the (lack of) robustness of edge modes to symmetry-respecting temporal fluctuations should hold generally.

We note that other topological aspects of non-equilibrium dynamics have been discussed in recent works. One direction concerns `Floquet-SPTs' \cite{Else2016,vonKeyserlingk2016,Potter2016,Roy2017}, wherein the periodic micromotion of periodically driven systems is characterized. A rather striking result in that context is the emergence of `anomalous' topologically protected edge modes which have no equilibrium analogue, in that their presence cannot be deduced from static properties of the Floquet Hamiltonian \cite{Rudner2013,Titum2016}. In a similar light, we demonstrate that a time-evolving wavefunction can possess robust topological properties even if the Hamiltonian which governs its dynamics is itself trivial, which further highlights the distinction between topology in and out of equilibrium. (The connections between Floquet-SPT order and the results of our work are discussed in Section \ref{secFloquet}.)
%In these studies, mechanisms to avoid chaotic heating must be put in place, and as such the systems in question must be either many-body localized (in low dimensions) or completely non-interacting. In contrast, our work reveals the topological structures embedded in instantaneous many-body wavefunctions driven under Hamiltonians with arbitrary time-dependence, which itself is robust to generic perturbations including weak interactions. Additionally, an important observation in this field the 

Separately, recent studies on non-interacting fermions \cite{Wang2017,Gong2017,Yang2018,Zhang2018} have demonstrated that static topological phases can be detected via dynamics. These characterizations of the full wavefunction trajectory after a quench, which can be measured via Bloch state tomography, are different from our non-equilibrium classification: They provide information about the Hamiltonian governing time evolution, rather than the wavefunction itself, and these protocols require translationally invariant, non-interacting systems.

Finally, there has recently been much attention focussed on the topological properties of systems which can be described with effective non-Hermitian Hamiltonians\cite{Esaki2011,Leykam2017,Shen2018,Gong2018}, which are intrinsically out of equilibrium. A complete treatment of the effects of non-Hermiticity on the topology of a time-evolving wavefunction is outside the scope of this work, however analogous results should be obtainable for such systems.

%The aim of this paper is to comprehensively characterize the topology of many-body wavefunctions which result from quenches of topological insulators and superconductors in arbitrary spatial dimension. In particular, we establish the existence of a
  %listing all possible topologically distinct non-equilibrium states as a function of spatial dimension and the symmetries of the initial and final Hamiltonians .

Our paper is structured as follows: In Section \ref{secSymmtry}, we review the concept of `dynamically-induced symmetry breaking' \cite{McGinley2018} -- the observation that unitary time evolution can break the symmetries of the wavefunction even if the Hamiltonian respects the symmetries at all times. We then discuss how pure non-equilibrium states can be topologically classified in Section \ref{secDefinition}, and go on to derive this classification for free fermion systems with non-spatial symmetries in Section \ref{secClassification}. Our classification is formulated such that it encapsulates all features familiar from equilibrium systems, which naturally leads to a non-equilibrium bulk-boundary correspondence that we develop in Section \ref{secEntanglement}. We then discuss direct physical consequences of our results in Section \ref{secPhysical}, before describing the relationships between our results and those found for Floquet systems in Section \ref{secFloquet}. We finally conclude in Section \ref{secConclusion}.

%Section \ref{secDefinition} then gives a precise definition of how one should define topological properties of a many-body wavefunction that is not an eigenstate of the governing Hamiltonian. This definition allows us to derive our classification table in terms of bulk properties of non-equilibrium states, which we construct in Section \ref{secClassification}. Section \ref{secEntanglement} offers a complementary perspective on the classification table, based on analysing bulk-boundary correspondences for individual states through the entanglement spectrum. Our results have direct physical consequences for experimental realisations of topological states, which we discuss in Section \ref{secPhysical}. We conclude in Section \ref{secConclusion}. 

\section{Symmetry of the time-evolved state \label{secSymmtry}}

We are concerned with the topological properties of many-body states after generic unitary time evolution. Of central importance to topological properties in general is the impact of symmetry constraints. These symmetries are imposed at the level of the microscopic Hamiltonian in question.
%For example, in the context of ultracold atomic gases in optical lattices \cite{Jaksch1998}, time-reversal symmetry is naturally present in the Hamiltonian due to the lack of coupling between the positional motion of neutral atoms and external magnetic fields. When studying systems in equilibrium, the symmetry of the Hamiltonian and the ground state is equivalent (assuming no spontaneous symmetry breaking).
However, the many-body states which we consider are far from equilibrium, and so we must be careful to distinguish symmetry properties of the post-quench state and the Hamiltonian. We have established this relationship in Ref.~\onlinecite{McGinley2018}; we review the results here.

The quench protocol which we will refer to in this paper is highly general: the system starts in the ground state of some initial Hamiltonian $\hat{\mathcal{H}}^\text{i}$ at time $t=0$ and evolves under some final Hamiltonian $\hat{\mathcal{H}}^\text{f}(t)$ which may itself vary in time. %Since the following sections are concerned with the topology of the state at some instant in time,
We then consider the properties of the state at some final time $t_\text{f}$.

If a symmetry is present in the initial Hamiltonian, then there exists some symmetry operator $\hat{\mathcal{O}}$ which commutes with $\hat{\mathcal{H}}^\text{i}$. By Wigner's theorem, the operator $\hat{\mathcal{O}}$ can be unitary or antiunitary, i.e.\@ $\hat{\mathcal{O}} i \hat{\mathcal{O}}^{-1} = \pm i$. Now, if the symmetry is not spontaneously broken, then the ground state of $\hat{\mathcal{H}}^\text{i}$ (which we call $\ket{\Psi_0}$) also respects this symmetry, in the sense that the pure density matrix $\hat{\varrho}_0 \coloneqq \ket{\Psi_0}\bra{\Psi_0}$ satisfies $[\hat{\mathcal{O}}, \hat{\varrho}_0]$. Note that we work with density matrices for convenience, but understand that they always represent a single pure wavefunction.

The state evolves as $\ket{\Psi(t)} = \hat{\mathcal{U}}(t)\ket{\Psi(0)}$, where $\hat{\mathcal{U}}(t) = \mathcal{T} \exp(-i\int_0^{t_\text{f}} \diff t' \hat{\mathcal{H}}^\text{f}(t'))$ is the time evolution operator ($\mathcal{T}$ denotes time-ordering). Thus the density matrix after the full quench is is $\hat{\varrho}(t_\text{f}) = \hat{\mathcal{U}}(t_\text{f}) \hat{\varrho}_0 \hat{\mathcal{U}}(t_\text{f})^\dagger$. To determine whether the symmetry $\hat{\mathcal{O}}$ is respected by the time-evolved state, we compute $[\hat{\mathcal{O}}, \hat{\varrho}(t_\text{f})]$, which will be zero if the symmetry is respected and non-zero otherwise.

If the final Hamiltonian $\hat{\mathcal{H}}^\text{f}(t)$ does not commute with $\hat{\mathcal{O}}$, then the state will also not respect the symmetry -- this we call explicit symmetry breaking. However, even if $\hat{\mathcal{H}}^\text{f}(t)$ satisfies the symmetry at all times, the state will not be symmetric ($[\hat{\mathcal{O}}, \hat{\varrho}(t_\text{f})] \neq 0$) if $\hat{\mathcal{O}}$ is an antiunitary operator -- this we call dynamically-induced symmetry breaking. One can see this as the factor of $i$ in the exponent of $\hat{\mathcal{U}}(t)$ is not invariant under antiunitary operations \cite{McGinley2018}.

It is therefore possible for the symmetry properties of the state and Hamiltonian to deviate out of equilibrium. This dynamically-induced symmetry breaking has profound consequences for topology out of equilibrium.

Although we do not do so here, one could also consider quench protocols in which the final Hamiltonian is non-Hermitian, which captures certain gain and/or loss processes. In doing so, one must be careful to consider the relationship between symmetries of the initial state and the non-Hermitian Hamiltonian, the latter of which can have a more general set of symmetries\cite{Lieu2018}. Note, however, that this work is concerned with properties of the instantaneous wavefunction rather than the spectrum of the Hamiltonian, in contrast to previous studies of non-Hermitian topology\cite{Esaki2011,Leykam2017,Shen2018,Gong2018}.

\section{Defining topology out of equilibrium \label{secDefinition}}

The topology of gapped short-ranged entangled (SRE) systems at zero temperature is a well-defined concept. Under a particular set of symmetry constraints, one constructs a set of topological \textit{phases}, defined such that Hamiltonians which belong to different phases cannot be continuously deformed between each other without breaking the symmetry constraints or crossing some topological quantum phase transition. In the majority of this paper, we are concerned with topological insulators under the `ten-fold way' (see Section \ref{secClassification}), where the scope of `continuous deformations' is rather broad. In particular, we are permitted to add trivial bands to a system, which for example excludes Hopf insulators \cite{Deng2013} from our definition.

This approach to equilibrium topology naturally gives rise to properties that are robust to perturbations including spatial disorder and weak interactions \cite{Niu1984,Xu2006}. In addition, such a definition (which is in terms of the bulk) correctly predicts the existence of gapless boundary edge modes. We show in this section that one can take a similar approach to classifying the topology of non-equilibrium states in a way that inherits all these important properties. These arguments apply to both noninteracting and interacting systems.

%In order to generalise the notion of topology to a post-quench wavefunction, we need to be precise in our definition of `topology'. Indeed the concept of equilibrium phases cannot be directly applied to time-evolved states. In this section we present a clear, physically motivated definition of topology out of equilibrium. Note that these arguments may not apply to long-ranged entangled phases that feature topological ground state degeneracies; such a study is outside the scope of this work.

\begin{figure*}
  	\includegraphics[scale=1.0]{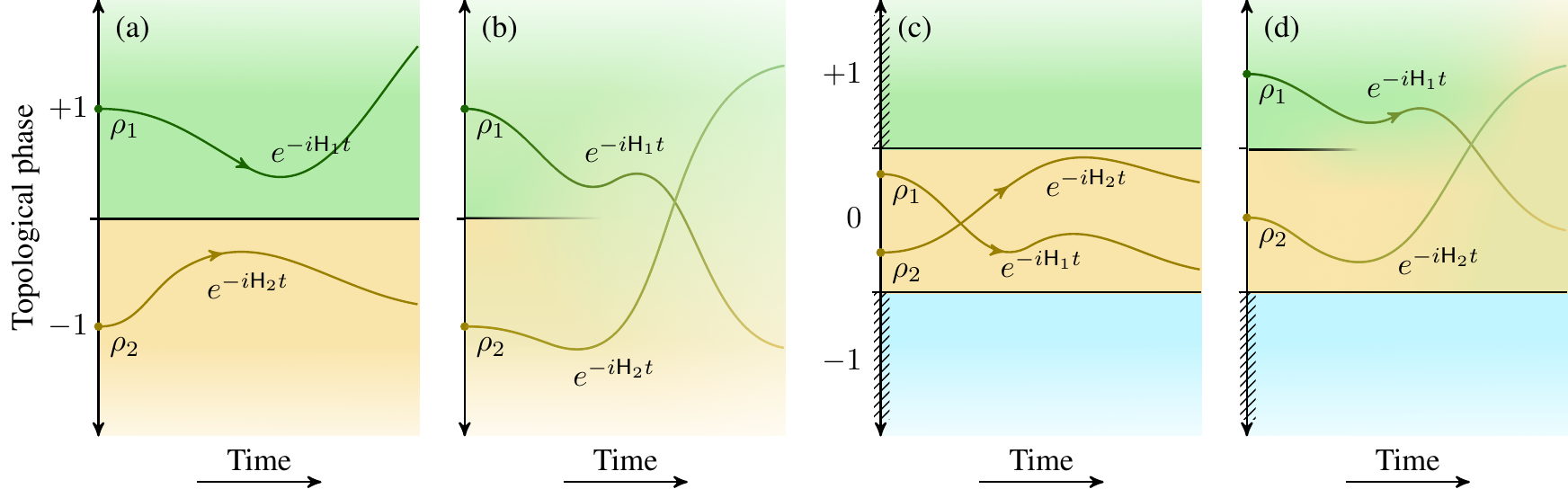}
  	\caption{Illustration of topological restrictions on unitary dynamics. The various topological phases are denoted `$0$', `$+1$', and `$-1$'. In each case (a--d), some phases are compatible with the symmetry constraints on the initial state, and others are not; the inaccessible phases are marked with hatches. Diagrams (a), (b), and (d) feature initial states $\matrrho_1$ and $\matrrho_2$ that are separated by an equilibrium topological phase transition (horizontal line), whilst in (c) the initial states are forced to be in the same equilibrium phase by the initial symmetries. In diagrams (a) and (c), the phase boundaries persist even after dynamically-induced symmetry breaking -- it is not possible for a state with one topology to evolve into a state with different topology. In (b) and (d), the topological distinction between the two phases is broken out of equilibrium, i.e\@ the initial state topology is lost under dynamics. In cases (b--d), the non-equilibrium classification is trivial due to initial state restrictions  and/or the breakdown of equilibrium phase boundaries due to symmetry breaking.}
  	\label{figTopology}
  \end{figure*}

Although equilibrium topology is commonly associated with physical observables, such as the Hall conductance, which manifestly depend on the full Hamiltonian, topological phases can be identified solely from the many-body ground state. This is evident for non-interacting fermions: Starting with a single-particle Hamiltonian $\hat{\mathcal{H}} = \hat{\psi}_i^\dagger \matr{H}_{ij} \hat{\psi}_j$ with single-particle energies $E_\lambda$, one can perform the spectral flattening procedure \cite{Schnyder2008,Ryu2010,Fidkowski2010}, where one smoothly modulates all the energies $E_\lambda$ to $\sgn E_\lambda = \pm 1$ without changing the eigenstates. Since this is a continuous deformation of the system which can be done without breaking any symmetry or closing the gap, the resultant Hamiltonian is topologically identified with the physical one. We arrive at the flat-band Hamiltonian $\matr{Q}$ which is related to the single-particle density matrix $\rho_{ij} = \braket{\Psi_0|\hat{\psi}_i^\dagger \hat{\psi}_j|\Psi_0}$ by $\matr{Q} = 1 - 2\matrrho$. This matrix $\matrrho$ is determined by the ground state of $\hat{\mathcal{H}}$ only. We see that all equilibrium topological properties are uniquely determined by two objects: 1) a single many-body state (the ground state), and 2) a collection of symmetries which define the allowed deformations of that state.

The above formulation of equilibrium topology has a natural non-equilibrium generalisation: We can use the time evolved state $\ket{\Psi(t_\text{f})}$ in place of the ground state, and use only the symmetries which are generically preserved under time evolution to define the allowed deformations. This is equivalent to constructing some fictitious Hamiltonian $\hat{\mathcal{Q}}(t_\text{f})$ for which $\ket{\Psi(t_\text{f})}$ is the ground state, and applying the equilibrium classification. The result is a topological classification which naturally inherits all the attractive physical features of the equilibrium classification.

%It is now clear how to generalise this definition to a non-equilibrium state $\ket{\Psi(t_\text{f})}$: The topology of the state is defined as the equilibrium topology of some fictitious Hamiltonian $\hat{\mathcal{Q}}$ for which $\ket{\Psi(t_\text{f})}$ is the ground state. Naturally, the symmetries which we use in turn to define the topology of $\hat{\mathcal{Q}}$ are those which are preserved under time evolution.

We will make use of the construction involving the fictitious Hamiltonian $\hat{\mathcal{Q}}(t_\text{f})$ in much of this paper. For concreteness (inspired by the spectral flattening procedure), we can use
\begin{align}
\hat{\mathcal{Q}}(t_\text{f}) = 1 - \ket{\Psi(t_\text{f})}\bra{\Psi(t_\text{f})}.
\label{eqHamiltonianProjector}
\end{align}
This construction can be applied to any gapped system, and at any finite time yields a Hamiltonian with couplings that decay exponentially with spatial separation \footnote{This follows from the fact that at some finite time the correlations of $\ket{\Psi(t_\text{f})}$, which become the matrix elements of $\hat{\mathcal{Q}}(t_\text{f})$ decay exponentially in space when sufficiently far apart.}.

Our definition of topology out of equilibrium is rather natural in that it pertains only to properties of $\ket{\Psi(t_\text{f})}$ (or equivalently $\hat{\mathcal{Q}}(t_\text{f})$) and the set of symmetries which the state generically possesses. We expect that our classification scheme will correctly predict all characteristic signatures of topology which are familiar from the study of systems at equilibrium. In particular, non-equilibrium states which are topologically non-trivial under our definition will exhibit a bulk-boundary correspondence, which can be seen through gapless edge modes of the entanglement spectrum \cite{Li2008}. We show this explicitly in Section \ref{secEntanglement}.
 
 With a less careful definition of non-equilibrium topology, one might expect that any two states that are initialised in topologically distinct wavefunctions must remain distinct for all time, since one can always evolve backwards in time using $\matr{U}(t)^\dagger$ and see that the two states began with different topologies. Such an approach was adopted in previous studies, such as Refs.~\onlinecite{Yang2018,Liu2018}. However, unlike our definition, this interpretation is a characterisation of the full historical trajectory of the wavefunction, since one cannot necessarily make the same conclusions from $\ket{\Psi(t_\text{f})}$ alone. We shall see that this definition would incorrectly predict the dynamics of the entanglement spectrum.
 
 Although the non-equilibrium classification we propose is set up in an analogous way to equilibrium, it can in general differ from the equilibrium classification due to the fact that not all symmetries are generically preserved under time evolution (see Section \ref{secSymmtry}). This means that the set of allowed deformations of the state in question is wider. Indeed, in equilibrium, two states are topologically equivalent if they can be smoothly deformed between one another without breaking the symmetries in question. However, out of equilibrium, we say that states are equivalent if they can be connected via unitary time evolution under some local Hamiltonian that respects the given symmetries. The former is a stronger statement than the latter.
 
 Furthermore, a na\"{i}ve expectation may be that the non-equilibrium classification will simply be given by the equilibrium classification under the subset of symmetries which are preserved dynamically. However, this neglects the fact that the initial state itself has some symmetry constraints, which may restrict which topological phases are accessible in the first place. As a simple example in the context of free-fermion systems (which are discussed in Section \ref{secClassification}), class AII systems in two spatial dimensions featuring time-reversal symmetry are reduced to having no symmetry (class A) once out of equilibrium. Although the equilibrium classification for this reduced symmetry class is the $\mathbb{Z}$-valued Chern number, the non-equilibrium classification will be trivial, since the requirement of a time-reversal invariant initial state necessitates the Chern number to start at zero, which is then preserved throughout evolution.
 
 We schematically illustrate various ways in which the classification can change out of equilibrium in Figure \ref{figTopology}. The four panels (a--d) represent possible outcomes for a given set of Hamiltonian symmetries (which we call `pre-quench symmetries'). After time evolution, the wavefunction respects only the `post-quench symmetries', which are the subset of pre-quench symmetries that are unitary. The possibilities we show are:
 
 %The remainder of this paper is concerned with establishing a topological classification for wavefunctions resulting from quenches between Hamiltonians with the same set of symmetries. The results of Section \ref{secSymmtry} demonstrate that the symmetry of the time-evolved state can be reduced by unitary dynamics, even if the symmetry is not explicitly broken in the Hamiltonian. We refer to the symmetries of the initial state as the `pre-quench symmetries', and the symmetries of the state after dynamically-induced symmetry breaking as the `post-quench symmetries'. %Clearly, the topology of the post-quench state $\ket{\Psi(t_\text{f})}$ must be classified under the post-quench symmetries.
 
 %A na\"{i}ve expectation may be that the classification of post-quench states is simply given by the equilibrium classification under the post-quench symmetries. However, the additional constraint of the pre-quench symmetries on the initial state may restrict the accessible topological phases out of equilibrium. Classifying non-equilibrium states involves relating the classifications of phases under the pre- and post-quench symmetries. We distinguish four possibilities for the fate of topology after a quench, and give examples of topological insulators which exhibit these behaviours (see Section \ref{secClassification}).
 {
 \renewcommand{\theenumi}{\alph{enumi}}
 \begin{enumerate}
 	\item The equilibrium classification under the pre-quench and post-quench symmetries is the same, and topologically distinct initial states ($\matrrho_1$ and $\matrrho_2$) remain topologically distinct for $t>0$. In this case, $\matrrho_1$ will not be able to evolve into a state with a different topology, and the `phase space' that can be explored under unitary dynamics is restricted, as illustrated in Figure \ref{figTopology}(a) (e.g.\@ Class CII in $d=4$).  
 	\item The equilibrium classification under post-quench symmetries is trivial. The topologically distinct initial states $\matrrho_1$ and $\matrrho_2$ become indistinct out of equilibrium, owing to some dynamically-induced symmetry breaking. Once the symmetry is broken for $t > 0$, the topological obstruction between the states is lifted, and there is no restriction on the accessible phase space, as illustrated in Figure \ref{figTopology}(b). (e.g.\@ Class AII in $d=3$).  
 	\item The equilibrium classification under post-quench symmetries is non-trivial, but the initial state classification is trivial. The additional symmetry restrictions on initial states ensures that all post-quench states are topologically equivalent, as illustrated in Figure  \ref{figTopology}(c). (e.g.\@ Class AI in $d=2$).  
 	\item The equilibrium classifications under pre- and post-quench symmetries are both non-trivial, but topologically distinct initial states are indistinct under the post-quench classification, as illustrated in Figure \ref{figTopology}(d).  (e.g.\@ Class DIII in $d=1$, or Class AII in $d = 2$).  
 \end{enumerate}
}

It is worth noting that here we are considering systems in the thermodynamic limit, or timescales which are subextensive in system size. For finite-sized systems, there will be a time on the order of $t \sim L/v_{\text{L.R.}}$ (where $v_\text{L.R.}$ is the Lieb-Robinson velocity) beyond which the correlations of $\ket{\Psi(t)}$ span the system size. At this point the system relaxes back into equilibrium at finite energy density, and our non-equilibrium classification no longer applies. Indeed such a thermalized state does not have a well-defined topology, since the fictitious Hamiltonian $\hat{\mathcal{Q}}(t_\text{f})$ will not be local. However, for times less than this thermalization time, the wavefunction topology is well-defined in the above sense. Additionally, if the system is many-body localized then topology remains well-defined for a time which grows exponentially with the system size.
 
 In the following section, we apply the formalism described above to systems of non-interacting fermions with non-spatial symmetries, yielding a topological classification that applies to non-equilibrium states.

\section{Classification of topological insulators out of equilibrium \label{secClassification}}

The `ten-fold way' of topological insulators \cite{Qi2008,Schnyder2008,Kitaev2009,Ryu2010} enumerates all topologically distinct phases of non-interacting fermionic systems subject to certain non-spatial symmetry constraints. Specifically, systems in a given spatial dimension $d$ belong to one of ten symmetry classes, depending on the presence of time-reversal (TRS), particle-hole (PHS), and chiral (or sublattice) symmetry \cite{Altland1997,Schnyder2008}. Each symmetry is represented by an on-site unitary matrix ($\matr{T}$, $\matr{C}$, or $\matr{S}$ respectively) subject to the constraints $\matr{T}\matr{T}^* = \pm 1$, $\matr{C}\matr{C}^* = \pm 1$ and $\matr{S}^2 = +1$ \cite{Chiu2016}. The entries of the ten-fold way are discrete groups (0, $\mathbb{Z}_2$ $\mathbb{Z}$, or $2\mathbb{Z}$), the elements of which represent different topological phases. The structure of the table is shown in Table \ref{tabEquilibrium} for reference. We shall construct a variant of this table which identifies the topological classes that are retained in the post-quench state, assuming that the initial and final Hamiltonians are in the same symmetry class. This is the aim of this section, the results of which are summarized in Table \ref{tabClassification}. Note that one could obtain analogous constructions for which the symmetry classes of the initial state and final Hamiltonians differ, however we do not do so here.

As described in Section \ref{secSymmtry}, the symmetry properties of the time-evolved state can differ from the symmetry class of the initial state and final Hamiltonian. Of the three symmetries featured in the ten-fold way, only PHS is unitary and thus due to dynamically-induced symmetry breaking, only PHS is preserved \cite{McGinley2018}. The post-quench symmetry class is thus either no symmetries (class A), PHS with $\matr{C}\matr{C}^* = +1$ (class D), or $\matr{C}\matr{C}^* = -1$ (class C). As we described in the previous section, the reduction of symmetry in the post-quench state can lead to a change of topology out of equilibrium.

\begin{table}
\begin{ruledtabular}
	\begin{tabular}{ll @{\hspace{0.35in}} ccc @{\hspace{0.1in}} c @{\hspace{0.25in}} c @{\hspace{0.05in}} }
		\multicolumn{2}{c}{Class} & \multicolumn{4}{c}{Symmetries} & Classification \\
		& & $\matr{T}$ & $\matr{C}$ & $\matr{S}$ & Label $s$ & $K_{\mathbb{C}, \mathbb{R}}(s,d=0)$ \\ \colrule
		Complex& A & 0 & 0 & 0 & $0$ & $\mathbb{Z}$  \\
		&AIII & 0 & 0 & 1 & $1$ & 0 \\ \colrule
		Real &AI & $+$ & 0 & 0 & $0$ & $\mathbb{Z}$ \\
		&BDI & $+$ & $+$ & 1 & $1$ & {$\mathbb{Z}_2$}  \\
		&D & 0 & $+$ & 0 & $2$ & {$\mathbb{Z}_2$}  \\
		&DIII & $-$ & $+$ & 1 & $3$ & 0  \\
		&AII & $-$ & 0 & 0 & $4$ & $2\mathbb{Z}$  \\
		&CII & $-$ & $-$ & 1 & $5$ & 0  \\
		&C & 0 & $-$ & 0 & $6$ & 0 \\
		&CI & $+$ & $-$ & 1 & $7$ & 0  \\
	\end{tabular}
\end{ruledtabular}
\caption{Periodic table of topological insulators at equilibrium \cite{Qi2008,Schnyder2008,Kitaev2009,Ryu2010}. The ten symmetry classes (two complex and eight real) are listed, and given a label $s$. The zero-dimensional classification is given, from which any dimensional classification can be inferred. For a given dimension $d$, the equilibrium classification is given by $K_\mathbb{C}(s - d \mod 2,0)$ in the complex classes and $K_\mathbb{R}(s - d \mod 8,0)$ in the real classes.}
\label{tabEquilibrium}
\end{table}

%Of the three available symmetries in the ten-fold way, PHS is the only one which is preserved. A na\"{i}ve approach to our problem would be to to assume that the non-equilibrium classification is simply given by the equilibrium classification for the corresponding symmetry class where TRS and chiral symmetries are absent (either class A, D, or C). However, this neglects the possibility that the symmetry of the initial state may further restrict the topology of the final state. As a simple example, a time-reversal symmetric $\mathbb{Z}_2$ topological insulator in 2D loses its TRS for $t > 0$, and accordingly the state belongs to class A, which is characterized by a $\mathbb{Z}$-valued Chern number. However, because the initial state possessed TRS, it must have been initialized with a zero Chern number, and is therefore necessarily trivial for $t > 0$. We have also identified more subtle cases in previous work \cite{McGinley2018}, such as class DIII in 1D.

To construct our classification, we will adopt similar approaches to those used to construct the equilibrium periodic table. In particular, the equilibrium classification can be obtained using the process of \textit{dimensional reduction} \cite{Qi2008}, in which the physical system is interpreted as a higher dimensional system with one (or more) of its dimensions compactified. This allows the topological properties of lower dimensional systems to be associated with `parent' systems in higher dimensions. Before generalizing to non-equilibrium states, we first review this procedure for systems in equilibrium, making reference only to properties of the ground state rather than the Hamiltonian. 

For simplicity of presentation, we describe strictly non-interacting translationally invariant systems. However, since we are concerned with strong topological invariants, we expect the results to hold in the presence of spatial disorder and weak interactions in a manner analogous to topological systems in equilibrium\cite{Niu1984,Xu2006}. Such arguments can be applied thanks to the interpretation of $\ket{\Psi(t_\text{f})}$ as the ground state of some local gapped Hamiltonian [Eq.~\eqref{eqHamiltonianProjector}] with non-trivial topology.
\begin{table*}
	\begin{ruledtabular}
		\begin{tabular}{l @{\hspace{0.25in}} ccc @{\hspace{0.45in}} cccccccc @{\hspace{0.05in}} }
			Class & \multicolumn{3}{c}{\hspace{-30pt}Symmetries} & \multicolumn{8}{c}{Spatial dimension $d$} \\
			& $\matr{T}$ & $\matr{C}$ & $\matr{S}$ & 0 & 1 & 2 & 3 & 4 & 5 & 6 & 7 \\ \colrule
			A & 0 & 0 & 0 & $\mathbb{Z}$ & 0 & $\mathbb{Z}$ & 0 & $\mathbb{Z}$ & 0  & $\mathbb{Z}$ & 0 \\
			AIII & 0 & 0 & 1 & 0 & \textcolor{oddprimary}{$\mathbb{Z} \rightarrow 0$}  & 0 & \textcolor{oddprimary}{$\mathbb{Z} \rightarrow 0$}  & 0 & \textcolor{oddprimary}{$\mathbb{Z} \rightarrow 0$}  & 0 & \textcolor{oddprimary}{$\mathbb{Z} \rightarrow 0$}\\ \colrule
			AI & $+$ & 0 & 0 & $\mathbb{Z}$ & 0 & 0 & 0 & \textcolor{even}{$2\mathbb{Z}$} & 0 & \textcolor{trsdesc}{$\mathbb{Z}_2 \rightarrow 0$} & \textcolor{trsdesc}{$\mathbb{Z}_2 \rightarrow 0$} \\
			BDI & $+$ & $+$ & 1 & \textcolor{odddescphs}{$\mathbb{Z}_2$} & \textcolor{oddprimary}{$\mathbb{Z} \rightarrow \mathbb{Z}_2$} & 0 & 0 & 0 & \textcolor{even}{$2\mathbb{Z} \rightarrow 0$} & 0 & \textcolor{odddescphs}{$\mathbb{Z}_2 \rightarrow 0$} \\
			D & 0 & $+$ & 0 & \textcolor{phsdesc}{$\mathbb{Z}_2$} & \textcolor{phsdesc}{$\mathbb{Z}_2$} & {$\mathbb{Z}$} & 0 & 0 & 0 & \textcolor{even}{$2\mathbb{Z}$} & 0 \\
			DIII & $-$ & $+$ & 1 & 0 & \textcolor{odddesctrs}{$\mathbb{Z}_2 \rightarrow 0$} & \textcolor{odddesctrs}{$\mathbb{Z}_2  \rightarrow 0$} & \textcolor{oddprimary}{$\mathbb{Z} \rightarrow 0$} & 0 & 0 & 0 & \textcolor{even}{$2\mathbb{Z} \rightarrow 0$} \\
			AII & $-$ & 0 & 0 & \textcolor{even}{$2\mathbb{Z}$} & 0 & \textcolor{trsdesc}{$\mathbb{Z}_2 \rightarrow 0$} & \textcolor{trsdesc}{$\mathbb{Z}_2 \rightarrow 0$} & {$\mathbb{Z}$} & 0 & 0 & 0  \\
			CII & $-$ & $-$ & 1 & 0 & \textcolor{even}{$2\mathbb{Z} \rightarrow 0$} &  0 & \textcolor{odddescphs}{$\mathbb{Z}_2 \rightarrow 0$} & \textcolor{odddescphs}{$\mathbb{Z}_2$} & \textcolor{oddprimary}{$\mathbb{Z} \rightarrow \mathbb{Z}_2$} & 0 & 0 \\
			C & 0 & $-$ & 0 & 0 & 0 & \textcolor{even}{$2\mathbb{Z}$} & 0 & \textcolor{phsdesc}{$\mathbb{Z}_2$} & \textcolor{phsdesc}{$\mathbb{Z}_2$} & $\mathbb{Z}$ & 0 \\
			CI & $+$ & $-$ & 1 & 0 & 0 & 0 & \textcolor{even}{$2\mathbb{Z} \rightarrow 0$} & 0 & \textcolor{odddesctrs}{$\mathbb{Z}_2  \rightarrow 0$} & \textcolor{odddesctrs}{$\mathbb{Z}_2  \rightarrow 0$} & \textcolor{oddprimary}{$\mathbb{Z} \rightarrow 0$} \\
		\end{tabular}
	\end{ruledtabular}
	\begin{tabular}{c @{\hspace{0.18in}}c @{\hspace{0.18in}}c @{\hspace{0.18in}}c @{\hspace{0.18in}}c}
		Even primary (\ref{secEvenPrimary}) & \textcolor{trsdesc}{Even descendants} (\ref{secEvenDesc})  & \textcolor{oddprimary}{Odd primary} (\ref{secOddPrimary})& \textcolor{odddesctrs}{Odd descendants} (\ref{secOddDesc}) & \textcolor{even}{$2\mathbb{Z}$ series} (\ref{sec2Z})
	\end{tabular}
	\caption{Classification of topological insulators out of equilibrium. The non-equilibrium classification describes the set of topological classes which remain distinct after time evolution under a Hamiltonian possessing the set of symmetries in question, as outlined in Section \ref{secDefinition}. The ten symmetry classes of the ten-fold way are listed on the left, and defined by the presence ($+$,$-$, $1$) or absence ($0$) of time-reversal ($\matr{T}$), particle-hole ($\matr{C}$), and chiral ($\matr{S}$) symmetries \cite{Altland1997,Schnyder2008}. For each symmetry class and spatial dimension $d$, the equilibrium and non-equilibrium classifications are given. A single entry indicates that the classification does not change out of equilibrium, whilst the notation $G_1 \rightarrow G_2$ indicates that the classification changes from $G_1$ in equilibrium to $G_2$ out of equilibrium. The different series of the classification are coloured as described in the main text, and the references to the discussions of each series are given below the table. Systems in dimension $d > 7$ have the same classification as the corresponding system in $(d - 8)$ dimensions (Bott periodicity).}
	\label{tabClassification}
\end{table*}

\subsection{Dimensional reduction in equilibrium}

\subsubsection{Non-chiral classes \label{secNonChiral}}

In the absence of symmetry (class A), systems in even dimensions $d = 2n$ can be characterized by the $n^\text{th}$ Chern number $\text{Ch}_n$. In terms of the Bloch wavefunctions $\ket{u^\alpha(\vec{k})}$ (where $\alpha$ labels the occupied bands), we define the non-Abelian Berry connection as a 1-form $\mathcal{A}^{\alpha \beta} = \braket{u^\alpha(\vec{k})|\diff u^\beta(\vec{k})}$, and the corresponding Berry curvature $\mathcal{F}^{\alpha \beta} = (\diff \mathcal{A} + \mathcal{A} \wedge \mathcal{A})^{\alpha \beta}$. Then the Chern numbers are given by an integral of the Chern form $\text{ch}_n$ over the Brillouin Zone (BZ) \cite{Ryu2010}
\begin{align}
\text{Ch}_n = \int_{\text{BZ}} \text{ch}_n\, , && \text{where}\; \text{ch}_n \coloneqq \frac{1}{(n+1)!}\Tr \left(\frac{i\mathcal{F}}{2\pi}\right)^{(n+1)}.
\end{align}

 Now consider adding either TRS (with $\matr{T}\matr{T}^* = \pm 1$) or PHS (with $\matr{C}\matr{C}^* = \pm 1$), yielding one of the non-chiral classes AI, AII, D, or C. Depending on the dimension and type of symmetry, this may or may not restrict the allowed values of $\text{Ch}_n$. If the Chern number is not restricted by the additional symmetry, then this entry in the periodic table is termed the `even primary series', and is $\mathbb{Z}$-classified \cite{Ryu2010}.

Each member of the even primary series induces two $\mathbb{Z}_2$ entries in two lower dimensions, termed the first and second descendants \cite{Qi2008}. For concreteness, we study the canonical example of the 4D primary insulator with TRS $\matr{T}\matr{T}^* = -1$, but all descendants can be understood in analogous ways.

Consider a 3D TRS state characterized by the density matrix $\matrrho^\text{(A)}(\vec{k})$ as our physical system. A one-parameter family of states $\matrrho(\vec{k},\theta)$ can be constructed which connects this insulator at $\theta = 0$ to some trivial TRS insulator $\matrrho^\text{ref}$ (which is independent of $\vec{k}$) at $\theta = \pi$. The intermediate 3D states for $0 < \theta < \pi$ need not possess TRS. Now consider closing this path into a loop $\theta = -\pi \rightarrow 0 \rightarrow \pi \equiv -\pi$ by invoking a `super-TRS' condition
\begin{align}
\matrrho(-\vec{k},-\theta) = \matr{T} \matrrho(\vec{k},\theta) \matr{T}^\dagger && 0 \leq \theta \leq \pi.
\label{eqSuperTRS}
\end{align}
Since $\matrrho^\text{(A)}(\vec{k})$ and $\matrrho^\text{ref}$ respect TRS, this loop can be made without any discontinuities. 

\begin{figure}
	\centering
	\includegraphics[scale=1]{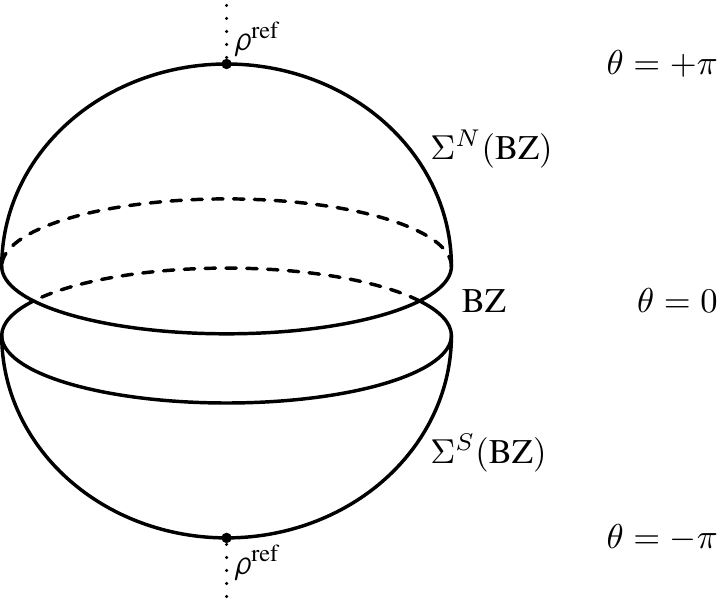}
	\caption{The physical Brillouin zone (BZ) as the equator of a higher dimensional momentum space $\Sigma(\text{BZ})$ parametrized by $(\vec{k},\theta)$. At the poles $\theta = \pm \pi$, the BZ is contracted to a point, representing the $\vec{k}$-independent reference state. We also identify the two poles, ensuring periodicity in $\theta$.}
	\label{figReductionTRS}
\end{figure}

By reinterpreting $\theta$ as an extra momentum variable in 4D, $\matrrho(\vec{k},\theta)$ represents a four dimensional TRS insulator, which is characterized by the second Chern number $\text{Ch}_2$. Because the reference Hamiltonian $\matrrho^\text{ref}$ is $\vec{k}$-independent, we can contract the subregions $\theta = \pm \pi$ to a single point, and so the higher dimensional momentum space is a `suspension' $\Sigma(\text{BZ})$, as illustrated in Figure \ref{figReductionTRS}.

Following Teo and Kane \cite{Teo2010}, one can show that the super-TRS condition \eqref{eqSuperTRS} forces the contributions to $\text{Ch}_2$ for $\theta > 0$ and $\theta < 0$ to be equal, and so we need only consider one hemisphere, which we call $\Sigma^N(\text{BZ})$. The Chern form $\text{ch}_2$ can be written as a total derivative of a 3-form called the Chern-Simons form $\text{ch}_2 = \diff Q_3$ \cite{Ryu2010}, and so the integral over $\theta > 0$ can be computed as a surface integral on the boundary $\theta = 0$, i.e.\@ the physical BZ. We then have
\begin{align}
\text{Ch}_2 = 2 \int_{\Sigma^N(\text{BZ})} \text{ch}_2 = 2 \int_\text{BZ} Q_3 \eqqcolon 2\text{CS}_3,
\label{eqHemisphereChern}
\end{align}
\noindent where $\text{CS}_3$ is the Chern-Simons (CS) invariant, which is entirely determined by the physical system at $\theta = 0$.

The CS invariant is gauge invariant only up to an integer. This gauge dependence reflects the fact that different embeddings of $\matrrho^\text{(A)}(\vec{k})$ in 4D can yield Chern numbers that differ by an even integer. $\text{Ch}_2 \mod 2$ defines a $\mathbb{Z}_2$-valued topological invariant which can characterize the 3D system unambiguously -- this relates the first descendant (3D) to the primary series (4D) in class AII. A similar construction is also possible for the second descendants, which are classified by the Fu-Kane (FK) invariant \cite{Fu2006}
\begin{align}
\text{FK}_{d = 2n} = \int_{\text{BZ}^{1/2}} \text{ch}_n - \int_{\partial \text{BZ}^{1/2}} Q_{2n-1},
\label{eqFK}
\end{align}
\noindent where $\text{BZ}^{1/2}$ is the half of the BZ where one of the momenta $0 \leq k_i < \pi$, and $\partial \text{BZ}^{1/2}$ is its boundary. To avoid ambiguity, this quantity must be calculated in a particular gauge that is specified by the TRS (or PHS) symmetry operator.

\subsubsection{Chiral classes \label{secChiral}}

Systems with only Chiral symmetry (class AIII) in odd dimensions also inherit their topology from a higher dimensional insulator in a similar way. The procedure is slightly different to the above, in that the higher dimensional insulator has a different symmetry to the physical state. Given a state $\matrrho^\text{(A)}(\vec{k})$ which respects a chiral symmetry operator $\matr{S}$, we can uniquely specify a higher-dimensional state via \cite{Teo2010b}
\begin{align}
\matrrho(\vec{k},\theta) &= \matrrho^\text{(A)}(\vec{k}) \cos (\theta/2) - \frac{1}{2}\matr{S} \sin (\theta/2) \nonumber\\ &+ \frac{1}{2}\left[1-\cos (\theta/2)\right] 
\qquad \theta \in [-\pi, \pi),
\label{eqChiralMapping}
\end{align}
\noindent where the last term enforces the correct trace. The addition of a term proportional to $\matr{S}$ breaks the chiral symmetry, and so the $(d+1)$-dimensional Hamiltonian belongs to class A and is characterizable by a Chern number. It can be shown \cite{Teo2010b} that topologically distinct chiral systems remain topologically distinct in the higher dimension, and vice-versa, i.e.\@ this mapping is a bijection between topological classes in different dimensions. Thus chiral systems in odd dimensions are $\mathbb{Z}$-classified according to the Chern number of $\matrrho(\vec{k},\theta)$. This can be more easily calculated as a winding number $\nu_{2n+1}$, or from the CS invariant calculated in a particular gauge \cite{Ryu2010}.

If TRS or PHS are also present in the physical system (classes BDI, DIII, CII, and CI), then we can still use the mapping \eqref{eqChiralMapping}. In this case the symmetry of the higher dimensional system changes according to [from $d$ to $(d+1)$ dimensions] \cite{Ryu2010}
\begin{align}
&\text{AIII} \rightarrow \text{A}; \hspace{10pt} \text{BDI} \rightarrow \text{D}; \hspace{10pt} \text{CII} \rightarrow \text{C}; \nonumber\\
&\text{DIII} \rightarrow \text{AII}; \hspace{10pt} \text{CI} \rightarrow \text{AI}.
\label{eqChiralReduction}
\end{align}
These relations are related to the ordering of the symmetry classes given in Tables \ref{tabEquilibrium} and \ref{tabClassification}. If the higher dimensional system belongs to the even primary series, then any Chern number is realisable and we say that the chiral system belongs to the odd primary series, which is also $\mathbb{Z}$-classified.

As in the non-chiral classes, the primary series gives rise to two descendants in the same symmetry class. As in Section \ref{secNonChiral}, a super-TRS condition is applied to the higher dimensional insulator, as well as chiral symmetry. In this case the fractional part of the CS invariant ($\text{CS}_{2n+1} \mod 1$) of the higher dimensional system determines the $\mathbb{Z}_2$ index for the physical system.

\subsection{Dimensional reduction out of equilibrium \label{secRedOOE}}

The above dimensional reduction procedure in equilibrium will motivate our approach to classifying non-equilibrium states. We generalize the method in the following way. Starting with an initial state $\matrrho(\vec{k},t=0)$ belonging to a particular symmetry class in $d$ dimensions, we construct the higher-dimensional insulator in $(d+r)$ dimensions $\matrrho(\vec{k},\vec{\theta},t=0)$ in the appropriate manner for the equilibrium classification. The physical ($d$-dimensional) system evolves under a final Hamiltonian $\matr{H}^\text{f}_{(d)}(\vec{k},t)$. We then dimensionally extend this final Hamiltonian to $\matr{H}^\text{f}_{(d+r)}(\vec{k},\vec{\theta},t)$ and consider the time-evolution of the higher-dimensional system, whilst ensuring that the $\vec{\theta} = \vec{0}$ subspace remains true to the physical system.

Of course, as in equilibrium, our conclusions regarding the topology of the physical system should be independent of the choice of embedding in this higher dimensional space; however one should be restricted to embeddings which respect the relevant symmetries of the system, e.g.~by enforcing the super-TRS condition \eqref{eqSuperTRS}. Furthermore, we are interested in those properties of the system that can be inferred from the instantaneous wavefunction $\ket{\Psi(t_\text{f})}$ alone, without reference to the history of the wavefunction at previous times $t < t_\text{f}$, as in Ref.~\onlinecite{McGinley2018}. If two choices of $\matr{H}^\text{f}_{(d)}(\vec{k},t)$ yield the same final state $\ket{\Psi(t_\text{f})}$, then our conclusions must be the same for both quench protocols. Crucially, when looking at $\ket{\Psi(t_\text{f})}$, one cannot distinguish whether TRS and chiral symmetries are broken dynamically or explicitly, since either process could yield the same final state. We therefore make no assumptions about the symmetry of $\matr{H}^\text{f}_{(d)}(\vec{k},t)$ except for the presence/absence of PHS, which can be inferred from $\ket{\Psi(t_\text{f})}$.

We note that, although in equilibrium the dimensional reduction parameter $\theta$ is often interpreted as a time coordinate which traces out an adiabatic evolution of the ground state, one should not confuse this parameter with the physical (generally non-adiabatic) time evolution in our non-equilibrium protocol. Instead, $\theta$ can be thought of as a coordinate which labels a one-parameter family of independent quench protocols.

Having described the general procedure, we now systematically construct our non-equilibrium table of topological insulators, considering each series in turn.

\subsubsection{Primary series in $d=2n$ \label{secEvenPrimary}}

As previously discussed, the primary series in even dimensions refers to the $\mathbb{Z}$-valued entries of the equilibrium table, and these systems are classified by the Chern number. Each member of the even primary series possesses one symmetry (TRS in $d = 4n$ or PHS in $d = 4n + 2$) which, heuristically, is irrelevant for the topology of the system, since the classification is neither restricted nor enriched by its addition. Indeed the Chern number must remain invariant under \textit{any} smooth gap-preserving deformations of the wavefunction, even if the underlying symmetries are broken.

Under unitary dynamics, the topology of the state is captured by the fictitious Hamiltonian $\hat{\mathcal{Q}}(t_\text{f})$ defined in \eqref{eqHamiltonianProjector}. The time $t$ parametrizes a smooth deformation connecting $\hat{\mathcal{Q}}(0)$ to $\hat{\mathcal{Q}}(t_\text{f})$, and therefore the Chern number (and hence the topology) of the initial state must be preserved in time. This behaviour has been proved for two dimensions in previous studies \cite{Foster2013,Sacramento2014,DAlessio2015,Caio2015,Toniolo2018}. Note that for finite systems, beyond a certain time $t \sim L/v_{\text{L.R.}}$ the correlations of $\ket{\Psi(t)}$ will span the whole system, at which point the Chern number is no longer a well-defined quantity. The primary series in $d=2n$ are coloured black in Table \ref{tabClassification}.

\subsubsection{First and second descendants in $d = 2n - 1$ and $d = 2n - 2$\label{secEvenDesc}}

The first and second descendants of the even primary series are constructed as described in Section \ref{secNonChiral}. Consider now the dynamics of the higher dimensional insulator.

If the descendants are PHS-protected (class D in $d = 0,1$ and class C in $d = 4,5$), then we can impose particle-hole symmetry on the dimensionally extended initial state $\matrrho(\vec{k},\vec{\theta},t=0)$ and final Hamiltonian $\matr{H}^\text{f}_{(d+r)}(\vec{k},\vec{\theta},t)$. This ensures that the PHS of the higher dimensional system is preserved in time, and so $\matrrho(\vec{k},\vec{\theta},t)$ will also respect PHS; the connection between insulators of different dimensions thus holds out of equilibrium. Moreover, the descendants inherit their topology from the even primary series, the topology of which is preserved (Section \ref{secEvenPrimary}). Therefore the topology of the descendants will not change in time.

However, if the descendants are TRS-protected (class AII in $d = 2,3$ and class AI in $d = 6,7$), then for $t > 0$ the $(d+r)$-dimensional state $\matrrho(\vec{k},\vec{\theta},t)$ will \textit{not} respect TRS due to dynamically-induced symmetry breaking. Even though the Chern number of $\matrrho(\vec{k},\vec{\theta},t)$ cannot change in time, the connection between the insulators of different dimension no longer holds. Indeed for first descendants, the relationship \eqref{eqHemisphereChern} between $\text{Ch}_n$ and $\text{CS}_{2n-1}$ no longer holds, because the contributions to $\text{Ch}_n$ for $\theta > 0$ and $\theta < 0$ are not equal once TRS is dynamically broken. Therefore the topology of TRS-protected first descendants is lost out of equilibrium.

For the first TRS descendants, we expect that the CS invariant will be free to vary continuously in time since there is no symmetry to quantize $\text{CS}_{2n-1}$ for $t>0$, in a similar way to class AIII in 1D \cite{McGinley2018}. This in turn implies that the second descendants, which themselves inherit their topology from the first descendants, must also lose their topology out of equilibrium. Unlike the first descendants, the relevant bulk index for second descendants [the FK invariant Eq.~\eqref{eqFK}], does not vary in time \cite{Liu2018}, however the above argument highlights that the relevance of the FK invariant to topology is lost when out of equilibrium. Indeed the FK invariant is only meaningful when a gauge determined by the TRS is adopted; once TRS is dynamically broken this gauge is no longer uniquely specified and thus the topology is lost. All the cases covered in this section are marked in blue in Table \ref{tabClassification}.

\subsubsection{Primary series in $d = (2n - 1)$ \label{secOddPrimary}}

We now turn to $\mathbb{Z}$-classified systems that feature a chiral symmetry, which constitute the odd primary series. In equilibrium, these systems are often analysed in terms of winding numbers \cite{Schnyder2008} without reference to dimensional reduction. However, once chiral symmetry is broken, the density matrix can no longer be brought into a canonical off-diagonal form and so the usual definition of the winding number is no longer well-defined. We will instead make reference to the dimensional reduction procedure outlined in Section \ref{secChiral}. Let us start with the case where no additional symmetries are present (class AIII in $d = 2n - 1$).

In equilibrium, the primary series in $d = (2n-1)$ can be related to the primary series in $(d+1)$ dimensions via the extension \eqref{eqChiralMapping}, and so we consider the time evolution of this $(d+1)$-dimensional insulator. As was highlighted at the beginning of Section \ref{secRedOOE}, we should not be able to distinguish explicit chiral symmetry breaking from dynamically-induced symmetry breaking, and so we should not assume that $\matr{H}^\text{f}_{(d)}(\vec{k},t)$ is chiral. Therefore we no longer have a unique prescription for dimensionally extending the final Hamiltonian, and so we can choose an arbitrary $\theta$-dependence for $\matr{H}^\text{f}_{(d+1)}(\vec{k},\theta,t)$.

\begin{figure}
	\includegraphics[scale=1]{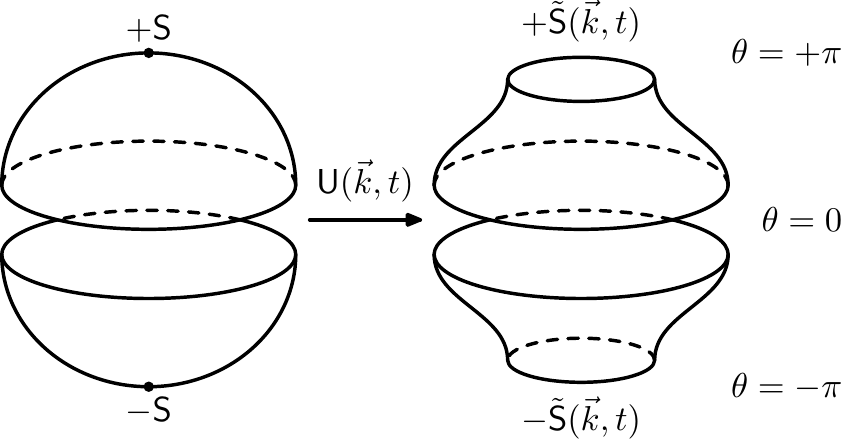}
	\caption{Dimensional extension of a chiral symmetric insulator after unitary time evolution. The higher dimensional BZ, which was compact at the poles $\theta = \pm \pi$, becomes open for $t > 0$.}
	\label{figChiralEvolution}
\end{figure}

Here we note an important difference between the dimensional extensions described in Sections \ref{secNonChiral} and \ref{secChiral}. In the former, the higher-dimensional insulator exhibits periodicity in the $\theta$ direction, due to the TRS/PHS of the reference system; this ensures that the BZ remains closed for all times, regardless of the choice of $\matr{H}^\text{f}_{(d+1)}(\vec{k},\theta,t)$. However, as can be seen from \eqref{eqChiralMapping}, the parent systems of chiral insulators are not periodic in $\theta$; rather, the closure of the BZ comes from the $\vec{k}$-independence of the state at the poles $\theta = \pm \pi$, which allows the poles to be compactified (see, e.g.\@ Ref.~\onlinecite{Teo2010b}). Since we are free to choose any embedding of $\matr{H}^\text{f}_{(d+1)}(\vec{k},\theta,t)$, the $\vec{k}$-independence at the poles may fail. Thus a boundary of the higher dimensional system may open up at $\theta = \pm \pi$, as illustrated in Figure \ref{figChiralEvolution}.

Because the BZ is no longer a closed manifold, the integral of the Chern form $\int_{\text{BZ}} \text{ch}_n$ is no longer quantized, as contributions can `leak' out through the boundary at $\theta = \pm \pi$. We could try to redefine a topological invariant by subtracting off the surface integral at the boundary, yielding a quantized index
\begin{align}
\mu(t)_\text{AIII} \coloneqq \int_{\text{BZ}_{(d+1)}} \text{ch}_n(t) - \int_{\partial \text{BZ}_{(d+1)}} Q_{2n-1}(t)
\label{eqIndexAIII}
\end{align}
\noindent where the surface $\partial \text{BZ}_{(d+1)}$ refers to the boundaries at $\theta = \pm \pi$. However, whilst $\mu(t)_\text{AIII}$ is quantized to an integer, its value is gauge dependent since the second term is gauge invariant only up to an integer. Therefore we cannot ascribe any physical meaning to $\mu(t)_\text{AIII}$. Clearly, the topology of class AIII systems is lost out of equilibrium.

For members of the primary series that possess TRS and PHS in addition to chiral symmetry, we must also consider the symmetry of the higher-dimensional system, which is given by \eqref{eqChiralReduction}. As before, the presence of TRS in the final Hamiltonian is irrelevant since the symmetry will be dynamically broken. However, if the higher dimensional system possesses PHS (as for BDI and CII), then we should also impose PHS on $\matr{H}^\text{f}_{(d+1)}(\vec{k},\theta,t)$, thereby preserving the symmetry of $\matrrho(\vec{k},\vec{\theta},t)$. As we found for the non-chiral classes, these PHS-respecting embeddings ensure that contributions to $\int_{\text{BZ}} \text{ch}_n$ are equal for $\theta < 0$ and $\theta > 0$, so we can only consider the upper half $\theta > 0$. Now we can construct an index analogous to that defined in \eqref{eqIndexAIII}
\begin{align}
\mu(t)_\text{PHS} \coloneqq 2\int_{\theta > 0} \text{ch}_n(t) - 2\int_{\theta = +\pi} Q_{2n-1}(t)
\label{eqIndexPHS}
\end{align}
This index is quantized to an integer, and owing to the factor of 2, gauge transforms can only change $\mu(t)_\text{PHS}$ by an \textit{even} integer. Thus the parity $\mu(t)_\text{PHS} \mod 2$ serves as a topological index which is preserved under unitary dynamics. Evidently, $\mu(t)_\text{PHS} \mod 2$ equals the parity of the $(d+1)$-Chern number at $t=0$. Additionally, if the first term of \eqref{eqIndexPHS} is evaluated using Stokes' theorem, then we find $\mu(t)_\text{PHS} = 2\text{CS}_{2n-1}(t)$. In essence, primary systems in classes BDI and CII are reduced to first descendants of even-dimensional systems, which have a $\mathbb{Z}_2$ classification. However for classes DIII and CI, the absence of PHS in the higher dimensional system results in the loss of topology for the same reasons as in class AIII. The odd primary series are marked in red in Table \ref{tabClassification}.

\subsubsection{First and second descendants in $d = (2n -1) - 1$ and $d = (2n - 1) - 2$\label{secOddDesc}}

$\mathbb{Z}_2$-classified insulators in the chiral classes inherit their topology from the odd primary series. Clearly, if the parent system loses its topology out of equilibrium (as is the case for classes DIII and CI), then its descendants will also lose their topology. 

On the other hand, in classes BDI and CII, the parent insulator is reduced from a $\mathbb{Z}$-classified primary insulator to the $\mathbb{Z}_2$-classified first descendant of the even primary series. We construct a higher-dimensional initial state in the same symmetry class according to Section \ref{secNonChiral}, and time-evolve under a PHS Hamiltonian. The $\mathbb{Z}_2$ classification of the descendant matches the \textit{parity} of the higher-dimensional winding number. Therefore, despite the reduction of the higher dimensional system from $\mathbb{Z}$ to $\mathbb{Z}_2$, the topology of the physical system is preserved, as it only depends on the bulk index modulo 2. We see that the first descendant of the odd primary series becomes the second descendant of the even primary series.

The second descendants of the odd primary series are $\mathbb{Z}_2$-classified in equilibrium; however generalizing the above construction would require us to reinterpret them as a \textit{third} descendant of the even primary series. As shown in \cite{Qi2008}, one cannot construct a third descendant with a non-trivial topological classification. Therefore, these systems lose their topology out of equilibrium. These systems studied in this section are marked in orange in Table \ref{tabClassification}.

\subsubsection{$2\mathbb{Z}$ classified systems\label{sec2Z}}

The only systems which remain to be classified are those which have a $2\mathbb{Z}$ classification in equilibrium; these occur four dimensions below the primary series. In even dimensions, these are classified by the Chern number just as in the primary series, but the extra symmetry enforces the Chern number to be even \cite{Ryu2010}. We can employ exactly the same reasoning as in Section \ref{secEvenPrimary} to show that these systems also preserve their topology out of equilibrium, with the caveat that systems must be initialised with a even Chern number.

Similarly, in odd dimensions the $2\mathbb{Z}$ systems are classified in the same way as the primary series, with the understanding that only even topological indices are possible. We use our results from Section \ref{secOddPrimary}, which show that in classes CI and DIII (in $d=3$ and $7$, respectively) the topology is lost. In classes CII and BDI (in $d=1$ and $5$, respectively), the parity of the winding number is preserved; however since the initial state must have an even winding number, all states become topologically trivial out of equilibrium. The systems covered in the above are coloured green in Table \ref{tabClassification}.

\subsection{Structure of the non-equilibrium classification}

Having considered all possible topological systems in all spatial dimensions, we arrive at our non-equilibrium classification given in Table \ref{tabClassification}. Some comments on its structure are required.

The fact that PHS is the only symmetry which is preserved under dynamics indicates that the state [or equivalently the Hamiltonian \eqref{eqHamiltonianProjector}] will collapse onto one of the symmetry classes A, D, or C. The equilibrium classification of these classes acts as a `upper bound', in that the non-equilibrium entry must be a subgroup of the corresponding equilibrium entry A, D, or C. For example, in $d=3$ and 7, the equilibrium classifications of classes A, D, and C are all 0, hence all non-equilibrium classifications in $d=3$ and 7 are 0.

The equilibrium table exhibits two forms of periodicity: Firstly, the table is invariant if all spatial dimensions are shifted by $d \rightarrow d+8$.  This is naturally also seen in our non-equilibrium table, since all our arguments are invariant under such an eightfold dimensional shift. The equilibrium table is also invariant if the dimension is increased by one and the symmetry classes are all shifted down. More precisely, the equilibrium classification only depends only on $s - d \mod 8\, (2)$, where $s$ is the label of the real (complex) symmetry class given in Table \ref{tabEquilibrium}. This full periodicity is not reflected in the non-equilibrium classification, because of the differences between the three symmetries: only PHS is preserved out of equilibrium. However, a subset of this periodicity survives. Between symmetry classes $s$ and $s+4$, all the symmetries are the same, with the exception that the quantities $\matr{T}\matr{T}^*$ and $\matr{C}\matr{C}^*$ change sign. Their role under dynamics is therefore the same, and so the non-equilibrium classification for $(s,d)$ is the same as for $(s+4,d+4)$.

\section{Bulk-boundary correspondence and entanglement spectra \label{secEntanglement}}

The arguments of Section \ref{secClassification} are formulated in terms of characterizations of the bulk of a system. This is a natural approach since bulk topological indices can be directly calculated for a single state composed of full Bloch bands. However, topological phases are also characterizable at their boundary, via the presence of certain gapless edge modes. We expect that one could alternatively derive our non-equilibrium classification by considering boundary modes, in a manner similar to the classification of edge theories in Ref.~\onlinecite{Schnyder2008}. Of course, a wavefunction alone does not itself possess edge excitations, since these are properties of a Hamiltonian spectrum, so these edge modes are not directly observable, but through our constructions described in Section \ref{secDefinition}, we can associate an edge theory to a state.

Specifically, since the topology of $\ket{\Psi(t_\text{f})}$ is given by the equilibrium topology of some fictitious Hamiltonian (e.g.\@ $\hat{\mathcal{Q}}(t_\text{f})$ in Eq.\@ \eqref{eqHamiltonianProjector}), we can consider the edge modes of this fictitious Hamiltonian. In the case of non-interacting fermions, we can calculate the time-evolved Bloch functions $\ket{u^\alpha(\vec{k},t)} = e^{-i\matr{H}^\text{f}(\vec{k})t}\ket{u^\alpha(\vec{k},0)}$ for an infinite system, and then construct a non-interacting translationally invariant Hamiltonian $\matr{Q}(\vec{k},t_\text{f})$ for which $\ket{\Psi(t_\text{f})}$ is the ground state. A real-space Hamiltonian $\matr{Q}(t_\text{f})$ can then be constructed from $\matr{Q}(\vec{k},t_\text{f})$, and given a boundary of codimension 1. If $\ket{\Psi(t_\text{f})}$ is topological, then gapless modes will appear at this edge.

This is a rather indirect way of probing the bulk-boundary correspondence associated with the topology of $\ket{\Psi(t_\text{f})}$, in part due to the fact that the choice of $\matr{Q}(t_\text{f})$ is not uniquely defined. A simpler strategy is to make use of the entanglement spectrum \cite{Li2008}, which is uniquely defined and can be calculated directly from the wavefunction $\ket{\Psi(t_\text{f})}$. In equilibrium, the entanglement spectrum reflects edge modes of the governing Hamiltonian, and can be computed from the ground state alone. By analogy, studying the entanglement spectrum of $\ket{\Psi(t_\text{f})}$ is a simple and direct way to detect the edge modes of $\hat{\mathcal{Q}}(t_\text{f})$.  To be specific, we partition the system into regions $A$ and $B$ and consider the reduced density matrix $\hat{\rho}_A = \Tr_B \ket{\Psi(t_\text{f})}\bra{\Psi(t_\text{f})}$. If the non-equilibrium state is topologically non-trivial, then we should see gapless entanglement modes in the entanglement Hamiltonian $\hat{\mathcal{H}}_E = -\ln \hat{\rho}_A$. Although there may be quantitative differences between the spectrum of $\hat{\mathcal{Q}}(t_\text{f})$ and the entanglement spectrum, the gaplessness of one spectrum implies that the other spectrum is gapless \cite{Fidkowski2010}.

 The entanglement spectrum is more computationally practical, and we will present numerical results for the entanglement spectrum in Section \ref{secEntNumerics}. However, since the two edge theories described above are equivalent, we simply refer to the fictitious Hamiltonian $\matr{Q}(t_\text{f})$ in our analysis, again considering cases where the initial and final Hamiltonians are in the same symmetry class.

\subsection{Edge theory analysis}

At $t = 0$, the Hamiltonian $\matr{Q}(t=0)$ will belong to the pre-quench symmetry class in question, and will possess an edge theory associated with the topological phase of $\matr{H}^\text{i}$. After time evolution, the Hamiltonian $\matr{Q}(t_\text{f})$ will belong to a symmetry class (A, D, or C), which may be reduced due to dynamically-induced symmetry breaking. Terms which were forbidden by TRS or chiral symmetry at $t=0$ may then appear in $\matr{Q}(t_\text{f})$. If these extra terms are able gap out the edge modes, then the boundary theory becomes trivial and topology is lost. However, if the edge modes survive for $t>0$, then topology is preserved.

In the following, we describe certain cases where edge modes are either preserved or destroyed; this analysis is not exhaustive, but it is clear how to generalise the arguments to arrive at Table \ref{tabClassification}. We neglect cases where the equilibrium classification of class A, D, or C is trivial, since in these cases $\matr{Q}(t_\text{f})$ will be unable to support any edge modes, regardless of the initial state.

\subsubsection{1D superconducting chains \label{secSupercond}}

The edge modes of one-dimensional systems are simply discrete modes at the zero-dimensional edges of the system. We compare classes BDI, D, and DIII, all of which reduce to class D for $t > 0$, and are realizable as topological superconductors \cite{Kitaev2001, Budich2013, Budich2013b}.

The edge states of these superconducting classes are all composed of Majorana fermions. In class D, the non-trivial phase simply hosts one Majorana fermion $\hat{\gamma}^{L(R)}$ at each edge $L$ or $R$. The addition of extra symmetries allows different types of Majorana to appear. Class BDI systems with a winding number $\nu$ will possess $|\nu|$ chiral Majorana zero-modes at each end $\hat{\gamma}_1^{L(R)}, \ldots, \hat{\gamma}_{|\nu|}^{L(R)}$, whereas non-trivial DIII systems possess a Kramers pair of Majorana fermions at each end $\hat{\gamma}_\text{I}^{L(R)}, \hat{\gamma}_\text{II}^{L(R)}$ \cite{Teo2010b} (here, Roman numerals label the two Kramers-degenerate states). After time evolution, chiral and TRS symmetries are broken, which allows local symmetry-breaking terms to appear in $\matr{Q}(t_\text{f})$. Majorana fermions on the same edge will be able to couple in pairs, but as long as PHS is preserved Majoranas on opposite ends cannot couple, and Majoranas cannot couple to the bulk \cite{Kitaev2001}. Any two Majoranas which do couple will become gapped. It is clear that in the DIII case, the existence of two Majoranas at each end means that this Kramers pair will in general become gapped, leading to a trivial edge theory, and hence a reduction of topology. On the other hand, in class BDI the $|\nu|$ Majoranas on a given edge will gap out in pairs. If $|\nu|$ is even, then all Majoranas will gap out and the edge theory will be trivial, but if $|\nu|$ is odd then one Majorana on each end will survive, corresponding to a non-trivial class D system. Therefore the $\mathbb{Z}$ classification of BDI is reduced to $\mathbb{Z}_2$, in agreement with Section \ref{secClassification}.

\subsubsection{2D insulators}

One can also make a similar analysis of the 1D edge modes of 2D systems. We contrast the chiral edge modes of class A systems with the TRS-protected helical modes of AII insulators \cite{Kane2005}. When $\text{Ch}_1 = +1$, the class A edge theory can be described using only one band, whilst the class AII edge features two bands (representing the spin degree of freedom). Taking a boundary perpendicular to the $x$ direction, the two edge theories can be written as
\begin{align}
\matr{H}^\text{b}_\text{A}(k_x) = v k_x; && \matr{H}^\text{b}_\text{AII}(k_x) = v k_x\, \sigma^z
\end{align}
\noindent where $\sigma^{x,y,z}$ are the Pauli matrices in spin space. The TRS operator in the AII case takes the form $\matr{T} = i\sigma^y$. The gapless nature of each edge theory is robust, in the sense that symmetry-respecting perturbations cannot open up a gap. For class A, this is simply due to the lack of other states to scatter into, whereas in class AII the TRS forbids any term that could open up a gap at $k_x = 0$.

\begin{figure}
\includegraphics[scale=1]{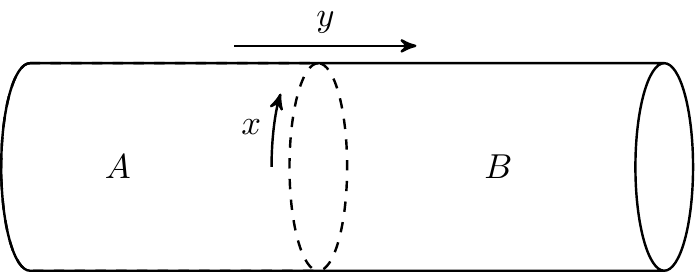}
\caption{Geometry of the entanglement cut for 2D systems with periodic boundary conditions in the $x$ direction and open boundary conditions with a large system size in the $y$ direction. The dashed line represents the divide between regions $A$ and $B$.}
\label{figCut}
\end{figure}

After a quench, the class A edge theory will remain gapless for the same reasons as in equilibrium; however for class AII, dynamically-induced symmetry breaking allows TRS-breaking terms to appear in the edge theory. For example, the term $m \sigma^x$, which is allowed after time evolution, will gap out the edge theory. Clearly, the $\mathbb{Z}_2$ edge mode is unstable under unitary dynamics, and the non-equilibrium topological classification can be identified as $\mathbb{Z}_2 \rightarrow 0$.

\begin{figure}

	\includegraphics[scale=1.0]{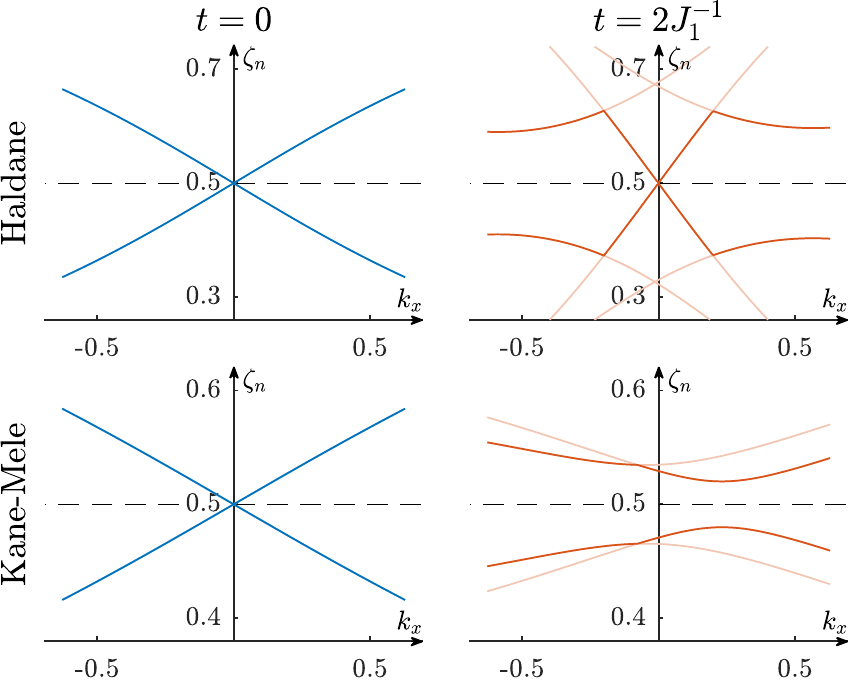}
	\caption{Dynamics of the single-particle modes of the entanglement spectrum (eigenvalues $\zeta_n$ of the reduced single-particle density matrix $C_{i,j} = \braket{\Psi(t)|\hat{\psi}_i^\dagger \hat{\psi}_j|\Psi(t)}$, where $i,j$ belong to the spatial region $A$) for the Haldane model (top) and the Kane-Mele model (bottom). In both systems, we start with a topologically non-trivial initial state at $t=0$ (left), and then time-evolve under a different Hamiltonian by a time $t = 2J_1^{-1}$. The entanglement spectrum of the time-evolved state is plotted (right). In the Haldane model, where topology is preserved, the edge state remains gapless after time evolution. However, in the Kane-Mele model, the gapless modes are TRS-protected and topology is destroyed out of equilibrium, hence the edge state becomes gapped at finite times.}
	\label{figEntanglement}
\end{figure}

\subsection{Numerical results for the entanglement spectrum\label{secEntNumerics}}

We supplement our analytical results on the dynamics of edge modes with some numerical simulations of the entanglement spectrum dynamics for 2D insulators. We use the Haldane model \cite{Haldane1988} and the Kane-Mele model \cite{Kane2005} as hosts of non-trivial class A and AII systems, respectively. We take periodic boundary conditions in the $x$ direction and make the entanglement cut perpendicular to the $y$ direction so that the wavevector $k_x$ is a good quantum number, as illustrated in Figure \ref{figCut}.

The Haldane model describes spinless fermions $\hat{c}_i^{(\dagger)}$ hopping on a honeycomb lattice (with sublattices $A$ and $B$), with Hamiltonian
\begin{align}
\hat{\mathcal{H}}_\text{Hal} &= J_1\sum_{\langle j, k\rangle} (\hat{c}_j^\dagger \hat{c}_k + \text{H.c.}) + J_2 \sum_{\llangle j, k \rrangle} \left(e^{i\phi_{jk}} \hat{c}_j^\dagger \hat{c}_k + \text{H.c.}\right) \nonumber\\ &+ m \sum_{j \in A} \hat{c}_j^\dagger \hat{c}_j - m \sum_{j \in B} \hat{c}_j^\dagger \hat{c}_j.
\end{align}
\noindent where $\langle j, k\rangle$ denotes nearest neighbours, and $\llangle j, k\rrangle$ denotes next-nearest neighbours. The phases $\phi_{jk}$ originate from a staggered magnetic flux, and are equal to $+\phi$ for anti-clockwise hopping about their common nearest neighbour, and $-\phi$ for clockwise hopping. The mass term $m$ serves to break the inversion symmetry of the lattice. The model possesses two bands corresponding to the sublattice degree of freedom, and realises Chern numbers of $0$, $+1$, and $-1$.

The Kane-Mele model has the same honeycomb structure, but features spinful fermions $\hat{c}_{i,\alpha}^{(\dagger)}$ where $\alpha = \uparrow, \downarrow$. Instead of a complex hopping (which breaks TRS), the model features a spin-orbit interaction as well as a Rashba interaction. The Hamiltonian is
\begin{align}
\hat{\mathcal{H}}_\text{KM} &= J_1\sum_{\langle j, k\rangle, \alpha} (\hat{c}_{j,\alpha}^\dagger \hat{c}_{k,\alpha} + \text{H.c.}) \nonumber\\ &+ i\eta_\text{s.o.} \hspace{-10pt} \sum_{\llangle j, k \rrangle, \alpha, \beta} \nu_{j,k} \hat{c}_{j,\alpha}^\dagger \sigma^z_{\alpha, \beta} \hat{c}_{k,\beta} \nonumber\\
&+ i\lambda_\text{R} \sum_{\langle j, k\rangle, \alpha, \beta} \hat{z} \cdot (\vec{\sigma} \times \vec{r}_{j,k})_{\alpha, \beta} \hat{c}_{j,\alpha}^\dagger \hat{c}_{k,\beta} \nonumber\\
&+m \sum_{j \in A, \alpha} \hat{c}_{j,\alpha}^\dagger \hat{c}_{j,\alpha} - m \sum_{j \in B, \alpha} \hat{c}_{j,\alpha}^\dagger \hat{c}_{j,\alpha}.
\end{align}
\noindent where $\nu_{j,k} = -1$ ($+1$) for clockwise (anti-clockwise) next-nearest neighbour hopping, and $\vec{r}_{j,k}$ is a unit vector in the direction from site $j$ to $k$. We have also included the inversion symmetry-breaking mass term.

To study the effect of non-equilibrium physics on the entanglement spectra, we construct an initial state as the ground state of the Hamiltonian in question. We then time-evolve under a final Hamiltonian which has different parameters, and look at the entanglement spectrum of the state after some finite time, which we choose to be $t_\text{f} = 2J_1^{-1}$ in both cases. For the Haldane model quench, we choose $(J_1, J_2, \phi, m)_{t=0} = (1,0.3,0.4,0.1)$, and then change the phase to $(\phi)_{t_\text{f}} = -0.2$. For the Kane-Mele model quench, we choose $(J_1, \eta_\text{s.o.}, \lambda_\text{R}, m)_{t=0} = (1,0.5,0.1,0.2)$, and then change the spin-orbit coupling to $(\eta_\text{s.o.})_{t_\text{f}} = 1.5$. Using the method of Peschel \cite{Peschel2003}, we obtain the entanglement spectrum by diagonalizing the `correlation matrix' (or the reduced single-particle density matrix) $\matr{C}_{i,j} = \braket{\Psi(t)|\hat{\psi}_i^\dagger \hat{\psi}_j|\Psi(t)}$. The eigenvalues of $\matr{C}$ are related to the single-particle excitation energies $\epsilon_n$ of the entanglement Hamiltonian $\hat{\mathcal{H}}_E$ via $\zeta_n = (e^{\epsilon_n}+1)^{-1}$. Therefore an eigenvalue equal to $\zeta = 0.5$ signals an entanglement degeneracy $\epsilon_n = 0$. The results are shown in Figure \ref{figEntanglement}. We see that the entanglement spectrum of the Haldane model remains gapless after the quench, however in the TRS-protected Kane-Mele model, the entanglement edge mode becomes gapped after the quench. This is consistent with our arguments of the previous section.

\section{Physical consequences \label{secPhysical}}

In the previous sections, we have used a number of different theoretical tools to better understand the topological structures of many-body wavefunctions out of equilibrium. Here we describe some consequences of our results that are directly relevant in experimental scenarios.

\subsection{Preparation and stability of topological states}

In an infinite 2D system, the time-independence of the Chern number implies that strictly a non-trivial Chern insulator ground state cannot be realised using unitary dynamics alone \cite{Caio2015,DAlessio2015}. Additionally, as mentioned in Section \ref{secDefinition}, even in a finite-sized system the Chern number remains constant until correlations span the system size, and the standard definition of the bulk invariant breaks down. Thus to realise a ground state of a topologically non-trivial Hamiltonian in cold atom experiments, one must adiabatically ramp the system across some topological phase transition \cite{Sorensen2010,Barkeshli2015}. Since the gap closes and the correlation length diverges at the phase transition, one must proceed slowly enough to avoid Landau-Zener tunnelling into an excited state. At the transition point, the gap to excited states is on the order of the level spacing $\sim (\Delta E)/N$, where $\Delta E$ is the band width of the Hamiltonian, and $N$ is the number of particles in the system. Thus to ensure the fidelity of adiabatic preparation, one must sweep across the transition over a time which grows extensively with the system size.

Our results generalise this observation to all classes of topological insulators and superconductors. Realising a topological state which has a non-trivial entry classification in Table \ref{tabClassification} cannot be achieved via unitary dynamics alone unless symmetries are explicitly broken in the governing Hamiltonian, or we proceed by slow adiabatic evolution. On the other hand, for systems which have trivial entries in Table \ref{tabClassification} it is possible to time-evolve from a trivial state to a topological one over a time which does not grow with the system size whilst respecting the symmetries of the Hamiltonian. Note that alternative non-adiabatic approaches to preparing topological states have recently been proposed which involve non-unitary dynamics, i.e.~interaction between the system and its environment \cite{Diehl2011,Budich2015}.

Conversely, topological states which are trivial under our non-equilibrium classification are generically unstable to time-dependent perturbations, for example external noise. In such systems, any fluctuations of the Hamiltonian with frequency component above the bulk gap will generically result in a state which has trivial topology. However, if the topological phase in question is stable out of equilibrium, then even in the presence of these fluctuations, the wavefunction will possess the same topology which it was initialised with.

Analogously, since in equilibrium bulk topological phases are intimately related with the protection of their edge modes, we expect that this instability of certain phases to time-dependent perturbations will also have important effects for the robustness of edge modes in general. In the following two sections, we study examples of these instabilities.

%We note that once the symmetry of a wavefunction is broken, regaining the symmetry at a later time through Hamiltonian engineering is a highly fine-tuned problem, and will in general only be possible for an instant in time. System-environment interactions can be employed to either remove the need for symmetry breaking, or as a means of restoring the symmetry of a state once it is broken.
%
% This instability can again be dealt with using dissipative dynamics. Intuitively, if system-environment interactions are stronger than time-dependent perturbations, then the system will remain in a topologically non-trivial steady state.

\subsection{Local adiabatic mixing of edge modes \label{secLocalMix}}

One proposed practical use of topological states is in quantum information technology \cite{Nayak2008,Sarma2015}. The non-local entanglement associated with 0D edge or defect modes allows for robust qubit storage over timescales that grow exponentially with system size -- indeed this was the original motivation of Kitaev's proposal to realise Majorana fermions \cite{Kitaev2001}. These edge modes are protected against static perturbations by the bulk topological phase in equilibrium. However, in cases where the bulk topological phase is destroyed out of equilibrium, we expect that the edge modes will not be protected against time-dependent perturbations. Here we discuss the way in which this dynamical instability affects 0D modes at the edges of 1D systems, however we expect that similar instabilities should arise for edge modes in higher dimensions, with equally important experimental consequences.

%Just as bulk topological phases can be symmetry-protected, the edge modes may rely on the presence of symmetries. For example, in systems which realise effectively spinless superconducting fermions (class D), the relevant symmetry is a PHS which emerges as a redundancy in the Bogoliubov-de Gennes (BdG) Hamiltonian \cite{Bernevig2013}. The Majorana edge modes which appear in the topological phase are therefore fully protected by the superconducting gap and the underlying structure of the BdG Hamiltonian. On the other hand, in one-dimensional time-reversal symmetric spinful topological superconductors (class DIII), the gapless nature of these modes must be protected by both the superconducting gap and the TRS imposed on the system (see Section \ref{secSupercond}).

 To facilitate quantum computation using 0D edge modes, one will need to be able to externally manipulate these topological qubits, which could be done through local adiabatic variation of some system parameters. The requirement of adiabaticity ensures that the qubit cannot couple to the bulk states, which are gapped out. However, the presence of Majorana edge modes implies the existence of a (almost) zero-energy subspace, which can be understood using degenerate adiabatic theory. Within this zero-energy subspace, there is no \textit{energetic} protection against transitions between different zero-energy modes when the Hamiltonian is varied in time, regardless of how slowly the variation is done. However, the modes may still be protected against mixing due to the symmetries that protect the bulk topological phase.
 
 From a practical perspective, we should distinguish mixing between Majoranas which is desired, and that which is undesired. The desired mixing will generally involve moving the Majoranas on a global scale, so that their non-trivial braiding statistics can be exploited \cite{Alicea2011,Liu2014}. The non-locality of this process ensures that it cannot happen `accidentally', i.e.~through a lack of control over the system. Conversely, any process which can happen locally is generally undesired, in that these processes could occur accidentally. Here we focus on the latter class of processes, and as such we will consider only a local part of the adiabatic low-energy subspace. Any mixing between edge modes within this local subspace is thus undesirable for quantum computation.

As the adiabatic process evolves, one must consider the full dynamics of this local low-energy subspace, which we assume is fully isolated from the bulk. Crucially, one should account for the possibility of dynamically-induced symmetry breaking within this subspace, and the associated reduction of topology described in the previous sections. Two edge modes which were protected against mixing by symmetry in equilibrium may be able to mix when the Hamiltonian is externally varied in time, because the symmetry which protects them is broken dynamically -- this indicates that such qubits would be susceptible to undesired local mixing. This local adiabatic mixing has been predicted in the specific case of class DIII in 1D \cite{Wolms2014}, and here we provide a framework by which this phenomenon can be understood more generally.

The question of which topological modes are vulnerable to local adibatic mixing is exactly equivalent to our previous consideration of which bulk topological phases are destroyed out of equilibrium, since we both problems reduce to the question of whether time-reversal and chiral symmetry-breaking terms are enough to lift the equilibrium topological protection. We therefore expect that our classification (Table \ref{tabClassification}) should generalize the results of \cite{Wolms2014} to all symmetry classes, in that entries which become trivial out of equilibrium indicate that the edge modes can adiabatically mix due to local perturbations, and therefore are inappropriate for qubit storage.

As an example, we consider 1D systems in class BDI which are in phases with winding number $|\nu| > 1$. Such systems would be expected to host $|\nu|$ chiral Majorana modes at an edge; however when non-equilibrium effects are considered, the topological classification reduces from $\nu_\text{eq.} \in \mathbb{Z}$ to $\nu_\text{non-eq.} \in \mathbb{Z}_2$. As a toy model of such a system (analogous to the one used to demonstrate mixing in class DIII in \cite{Wolms2014}), we use a semi-infinite extended Kitaev chain with beyond-nearest-neighbour hopping and pairing \cite{Niu2012,Lieu2018}

\begin{align}
\hat{\mathcal{H}}_{\nu>0} = \frac{E_g}{2}\sum_{j=1}^{\infty} i\hat{\gamma}_j^B \hat{\gamma}_{j+\nu}^A.
\label{eqHamIdeal}
\end{align}
Here, we define Majorana operators $\hat{\gamma}^A_j = \hat{c}_j + \hat{c}_j^\dagger$ and $\hat{\gamma}_j^B = -i(\hat{c}_j - \hat{c}_j^\dagger)$ in terms of the spinless fermion creation and annihilation operators $\hat{c}_j,\hat{c}_j^\dagger$. The Hamiltonian features equal amplitude hopping and $p$-wave superconducting pairing. Clearly, the Majorana modes $\hat{\gamma}_1^A \cdots \hat{\gamma}_\nu^A$ are not involved in the Hamiltonian and hence constitute a local set of zero-energy modes. These modes cannot be gapped out as long as the time-reversal symmetry (associated with the realness of the hopping and p-wave pairing) is preserved. Note that this time reversal symmetry ensures that terms with an even number of `$A$' or `$B$' labels are not allowed, e.g.~$\hat{\gamma}_j^A\hat{\gamma}_k^A$ is forbidden.

Class BDI in $d=1$ has a non-equilibrium classification $\mathbb{Z} \rightarrow \mathbb{Z}_2$, and so when $\nu = 2$ we expect the topological protection of edge modes to be lifted. For this value of $\nu$, we only need to consider six Majorana operators: $\hat{\gamma}_{1,2,3,4}^A$ and $\hat{\gamma}_{1,2}^B$, since all other operators decouple. We consider gradually turning on additional terms in the Hamiltonian which satisfy all the required symmetries, and act only on these six local Majoranas. Two such terms are
\begin{align}
\hat{\mathcal{H}}_\mu &= \frac{i\mu}{2}\left[\hat{\gamma}_1^A\hat{\gamma}_1^B + \hat{\gamma}_2^A\hat{\gamma}_2^B\right] \nonumber\\
\hat{\mathcal{H}}_J &= \frac{iJ}{2}\left[\hat{\gamma}_1^A\hat{\gamma}_2^B + \hat{\gamma}_2^A\hat{\gamma}_1^B\right].
\label{eqBDIPert}
\end{align}
The first applies a chemical potential to the first two sites, whilst the second enhances the single-particle hopping between the first two sites. Following Ref.~\onlinecite{Wolms2014}, we calculate the degree of mixing between the two modes $\hat{\gamma}_1^A , \hat{\gamma}_2^A$ using a non-Abelian Berry connection. Since the variation is slow with respect to $E_g$, a Majorana zero mode $\Gamma$ (which squares to 1) must remain an instantaneous zero-energy eigenstate after time evolution, which means we must have $\Gamma(t) = \cos\varphi(t) \hat{\gamma}^\text{I}_{\vec{\eta}(t)} + \sin\varphi(t) \hat{\gamma}^\text{II}_{\vec{\eta}(t)}$, where the Roman numerals distinguish the two instantaneous zero-modes. The dynamics of the mixing angle $\varphi(t)$ follow
\begin{align}
\varphi(t) = \varphi(0) + \int_{\eta(0)}^{\eta(t)} \diff \vec{\eta} \cdot \mathcal{A}(\vec \eta),
\end{align}
where the Berry connection is
\begin{align}
\mathcal{A}(\vec \eta) = \frac{1}{2}\left\lbrace \hat{\gamma}^\text{I}_{\vec{\eta}}, \vec{\nabla}_\eta \hat{\gamma}^\text{II}_{\vec{\eta}} \right\rbrace
\end{align}
with associated Berry curvature $\Omega_{ij} = \partial_{\eta_i}\mathcal{A}_{\eta_j} - \partial_{\eta_j}\mathcal{A}_{\eta_i}$. For the model considered above, we find a non-zero Berry curvature $\Omega_{\mu J} = -E_g^2(E_g^2 + \mu^2 + J^2)^{-2}$ \footnote{Despite the difference in symmetries of the two models, there is in fact a formal mapping between the six-Majorana system considered here and the one considered in Ref.~\onlinecite{Wolms2014}, which means the Berry connections are equal if one replaces $J$ in {\protect \eqref{eqBDIPert}} with $D$ in Eq.~(6b) of \cite{Wolms2014}. }, indicating that in general there will be a non-zero amplitude for one Majorana mode to evolve into the other.

\subsection{Decoherence of Majorana-based qubit storage due to noise}

The local adiabatic mixing described above demonstrates how temporal variation of external parameters can lead to mixing between degrees of freedom that are topologically protected in equilibrium. In the context of quantum computation, such external variation is necessary for manipulating and/or accessing the information held in the Majorana qubits. However, even when the qubit is simply stored and not accessed, local fluctuations due to external noise may be present. These fluctuations, although non-deterministic, can still lead to mixing between the local degrees of freedom if the topological phase in question is unstable in the non-equilibrium classification (Table \ref{tabClassification}). When the noise is accounted for and averaged over, this should result in decoherence of the qubit which was initially stored. In this section, we show that this decoherence can even occur when the noise is statistically time-reversal symmetric, so that there is no external bias between forward and backward time directions in the noise correlation functions. Note that, due to the adiabatic nature of the mixing, this decoherence should appear for arbitrarily small noise frequencies -- specifically we show later that the decoherence time to scale with the noise frequency $\Gamma$ as $\tau_\text{d} \sim E_\text{gap}^2 V^{-2} \Gamma^{-1}$, where $V$ is the amplitude of the noise. This is in stark contrast to decoherence due to mixing of edge and bulk states, mediated by noise with frequency components above the gap \cite{Goldstein2011,Moller2011,Schmidt2012}.

To test our hypothesis that the non-equilibrium classification predicts which topological zero-modes are unstable to this decoherence, we numerically simulate systems in classes BDI and DIII in $d=1$ which are subject to external noise, and determine the extent to which information stored in the Majorana modes is lost in time. Moreover, since class BDI has an entry $\mathbb{Z}$ (equilibrium) $\rightarrow \mathbb{Z}_2$ (non-equilibrium), we wish to compare systems in that class with different parities of the topological index, thus we take the winding number $\nu$ to be 1 (stable) and 2 (unstable). The three models we consider, therefore, are
\begin{widetext}
\begin{subequations}
\begin{align}
\hat{\mathcal{H}}^\text{DIII} &= \sum_{j,\sigma}\left[\vphantom{\sum}\frac{1}{2}\mu_j \hat{c}_{j\sigma}^\dagger\hat{c}_{j\sigma} +  J_j \hat{c}_{j\sigma}^\dagger\hat{c}_{j+1\,\sigma} + \Delta^{}_j \hat{c}_{j\sigma}^\dagger\hat{c}_{j+1\,\sigma}^\dagger \right] 
%\nonumber\\ &
+ \sum_{j}\left[ \vphantom{\sum} \Delta^{(s)}_j \hat{c}_{j\uparrow}^\dagger\hat{c}_{j\downarrow}^\dagger + \alpha^R_j \left(\hat{c}_{j\uparrow}^\dagger \hat{c}_{j+1\downarrow} + \hat{c}_{j\downarrow}^\dagger \hat{c}_{j+1\uparrow}\right) \right]+ \text{h.c.};\label{eqDecoMdoelA} \\
\hat{\mathcal{H}}^\text{BDI}_{\nu = 1} &= \sum_{j,\beta} \left[\vphantom{\sum}\frac{1}{2}\mu_{j\beta} \hat{c}_{j\beta}^\dagger\hat{c}_{j\beta} +  J_{j\beta} \hat{c}_{j\beta}^\dagger\hat{c}_{j+1\,\beta} + \Delta_{j\beta} \hat{c}_{j\beta}^\dagger\hat{c}_{j+1\,\beta}^\dagger\right]+ \text{h.c.}; \label{eqDecoMdoelB} \\
\hat{\mathcal{H}}^\text{BDI}_{\nu = 2} &= \sum_{j} \left[\vphantom{\sum}\frac{1}{2}\mu_j \hat{c}_{j}^\dagger\hat{c}_{j} +  J_j^{} \hat{c}_{j}^\dagger\hat{c}_{j+1} + \Delta^{}_j \hat{c}_{j}^\dagger\hat{c}_{j+1}^\dagger + J_j^{(2)} \hat{c}_{j}^\dagger\hat{c}_{j+2} + \Delta^{(2)}_j \hat{c}_{j}^\dagger\hat{c}_{j+2}^\dagger\right] + \text{h.c.}\;. \label{eqDecoMdoelC}
\end{align}
\label{eqModels}
\end{subequations}
\end{widetext}
Model \eqref{eqDecoMdoelA} features fermions $\hat{c}_{j\sigma}$ with a spin-1/2 index $\sigma$; model \eqref{eqDecoMdoelB} features fermions $\hat{c}_{j\beta}$ where the label $\beta = 1,2$ distinguishes two disconnected chains; and Model \eqref{eqDecoMdoelC}, which generalizes the fine-tuned Hamiltonian \eqref{eqHamIdeal} to include generic terms allowed by symmetry, features spinless fermions $\hat{c}_{j}$. The various terms featured in the models, all of which can vary spatially, are a chemical potential $\mu_j$; a Rashba spin-orbit coupling term $\alpha^R_j$; a single-particle hopping amplitude $J_j$; and a $p$-wave ($s$-wave) superconducting pairing amplitude $\Delta$ ($\Delta^{(s)}$). The $p$-wave superconducting and hopping amplitudes can couple fermions either 1 or two sites apart -- this difference allows us to access both the $\nu = 1$ and $\nu = 2$ phases in the class BDI cases. Each single-particle Hamiltonian will respect PHS ($\matr{C}\matr{C}^* = +1$) due to the redundancy of the Bogoliubov-de Gennes description \cite{Bernevig2013}. In addition, when the parameters are real, both systems satisfy a TRS. In the spinful system \eqref{eqDecoMdoelA} the TRS is symplectic ($\matr{T}\matr{T}^* = -1$), putting it in class DIII. On the other hand, the latter two models possess a TRS satisfying $\matr{T}\matr{T}^* = +1$ due to the spinless nature of the fermions, and hence belong to class BDI.

The chains are duplicated in the model of \eqref{eqDecoMdoelB} so that each of the three systems possesses 4 Majorana zero modes, which is the minimum number required to store a qubit without violating the fermion parity superselection rule. Therefore, each model possesses two `left' $\hat{\gamma}^{1,2}_\text{L}$ and two `right' $\hat{\gamma}^{1,2}_\text{R}$ Majorana zero modes, which in models \eqref{eqDecoMdoelA} and \eqref{eqDecoMdoelC} are protected against being gapped out by a time-reversal symmetry (equivalently, a chiral symmetry).

In each case, the low energy subspace consists of four states for which the bulk is in its ground state, and the non-local Dirac fermions $\hat{a}^\alpha = \hat{\gamma}^\alpha_\text{L} + i\hat{\gamma}^\alpha_\text{R}$ ($\alpha = 1,2$) are occupied or unoccupied. The fermion parity sectors cannot mix, and so for concreteness we consider only the parity sector where an odd number of edge modes are occupied ($\hat{\gamma}_\text{L}^1\hat{\gamma}_\text{R}^1\hat{\gamma}_\text{L}^2\hat{\gamma}_\text{R}^2 =+1$) so that the basis states for the qubit are the states $\ket{1,0}$ and $\ket{0,1}$, where $\ket{n^1,n^2}$ denotes the states with $\braket{\hat{f}^{\alpha\,\dagger} \hat{f}^\alpha} = n^\alpha$. One can use Pauli operators $\hat{\sigma}^z = i\hat{\gamma}_\text{L}^1\hat{\gamma}_\text{R}^1; \hat{\sigma}^x = i\hat{\gamma}_\text{L}^1\hat{\gamma}_\text{L}^2$ as a basis of operators on this qubit space.

As would be expected in practice, the noise which we introduce in these systems is local. Since we are interested in the response of the edge modes, we choose noise sources which act only on the two leftmost and two rightmost sites in each system independently. At each end, we consider two simultaneous noise terms which are mutually uncorrelated, but overlap spatially. (The reason for considering two noise sources is discussed below.)  For each term, the time dependence of the parameter $\eta(t)$ in question is an independent random signal which is zero at $t=0$; has mean equal to its initial value; and with a Lorentzian power spectrum, i.e.~the noise correlator $C(t) \coloneqq \noiseavg{\eta(t')\eta(t'+t)} - \noiseavg{\eta(t')}^2$ has a Fourier transform $\tilde{C}(\omega) \propto (\omega^2 + \Gamma^2)^{-1}$, where the width $\Gamma$ characterizes the noise frequency. Note that the noise we consider is statistically time-reversal symmetric, in the sense that $C(t) = C(-t)$.

To quantify the loss of information due to dephasing, we use the `recovery fidelity' developed by the authors of Ref.~\onlinecite{Mazza2013}, wherein the robustness of class D Majorana-based memories to global fluctuations was studied. This quantity characterizes the extent to which the initial information stored can be recovered by some optimal recovery process. To calculate the fidelity, the authors consider two initial pure qubit states, labelled by $+,-$ which are opposite on the Bloch sphere, i.e.~states such that the density matrix in the Majorana subspace is $\hat{\rho}_\pm = \hat{\rho}^\text{Bulk}_0 \otimes \hat(1\pm \hat{\sigma}^x)/2$, where $\hat{\rho}^\text{Bulk}_0$ is the ground state density matrix of the bulk, and the $\hat{\sigma}^x$ acts in the Majorana subspace. These initial states are then evolved for a time $t$ under the same realization of the noise potential, and the states obtained from different realizations are averaged to obtain mixed density matrices $\hat{\rho}_\pm(t)$. Ref.~\onlinecite{Mazza2013} showed that the optimal Gaussian recovery process has a fidelity
\begin{align}
F^\text{opt}(t) = \frac{2}{3} + \frac{1}{6}\left\|\Gamma_+(t) - \Gamma_-(t)\right\|_{\text{op}}
\label{eqFidelity}
\end{align}
Here, $\| \cdot \|_{\text{op}}$ is the operator norm (which returns the largest eigenvalue), and $\Gamma_\pm(t)_{jk} \coloneqq \Tr [\hat{\rho}_\pm (t) \hat{\gamma}_j \hat{\gamma}_k ]$ is the covariance matrix, where $j,k$ label the set of Majorana operators in the system.

\begin{figure}
	\includegraphics[scale=1]{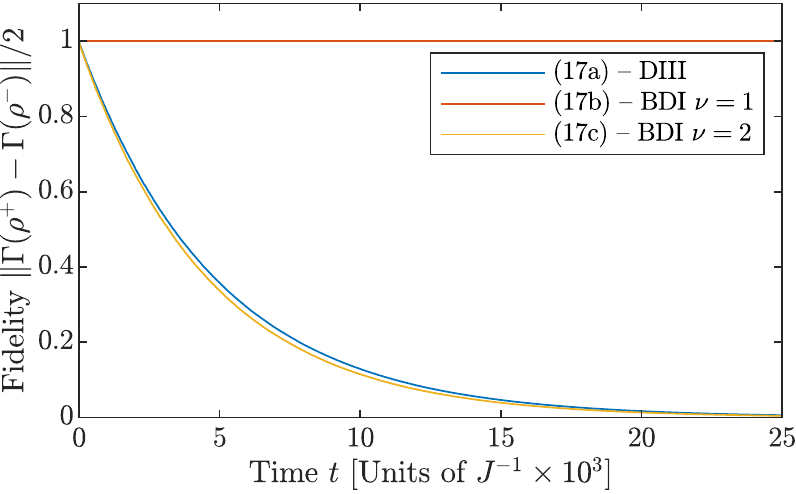}
	\caption{Decoherence of Majorana qubit memories due to temporal noise, as witnessed by the recovery fidelity [Eq.~\eqref{eqFidelity}]. We compare three systems in $d=1$ [with Hamiltonians given in Eq.~\eqref{eqModels}], in symmetry classes DIII and BDI -- these have entries $\mathbb{Z}_2 \rightarrow 0$ and $\mathbb{Z} \rightarrow \mathbb{Z}_2$, respectively, in the non-equilibrium classification (Table \ref{tabClassification}). Accordingly, the topology of models \eqref{eqDecoMdoelA} and \eqref{eqDecoMdoelC} is unstable out of equilibrium. This is reflected in the fidelity of storage in the associated Majorana modes when local Lorentzian (TRS) noise is present: The fidelity for the stable model \eqref{eqDecoMdoelB} saturates at a constant value, indicating the preservation of the qubit, whereas the initial state information stored in the unstable models decays away completely, indicating that there is no measurement which can be made to extract the initial qubit state.}
	\label{figDecoherence}
\end{figure}

We calculate the time dependence of the fidelity for each of the models \eqref{eqModels} and plot the results in Figure \ref{figDecoherence}. All Hamiltonian parameters are site-independent, except for the noise terms acting on the two leftmost sites. The initial Hamiltonian parameters chosen are: $(\mu, J, \Delta^{}, \Delta^{(s)}, \alpha^R)^\text{DIII} = (0.25,1,1,0.3,0.2)$ in \eqref{eqDecoMdoelA}; $(\mu, J, \Delta)^\text{BDI}_{\nu = 1} = (0.25,1,1)$ in \eqref{eqDecoMdoelB}; and $(\mu, J^{}, J^{(2)} \Delta^{}, \Delta^{(2)})^\text{BDI}_{\nu = 2} = (0.25,0.2,1,0.3,1)$. These values are chosen such that the systems are all in the desired phases, and have approximately equal decay lengths for the Majorana wavefunctions. Noise is introduced at each edge through an explicit time dependence of $\mu_j$ and $\Delta^{(s)}_j$ in \eqref{eqDecoMdoelA}; $\mu_{j\, \beta=1,2}$ and $\Delta_j$ in \eqref{eqDecoMdoelB}; and $\mu_{j}$ and $\Delta_j^{}$ in \eqref{eqDecoMdoelC} (with $j = 1,2$ on the left edge, and $j = N-1,N$ on the right edge). In $\hat{\mathcal{H}}_{\nu = 1}^\text{BDI}$, the noise signals on the two disconnected wires are independent and uncorrelated. These noise signals have an amplitude such that the root mean square of the signal $\sqrt{C(t=0)} = 0.1$. We choose the width of the Lorentzian noise spectrum to be small $\Gamma = 0.02$, so as to minimise coupling of the edges and bulk. The length of each chain is $N = 24$, and the density matrices are averaged over 20 noise realisations.

In model \eqref{eqDecoMdoelB}, the bulk topology is preserved under the noise, since the topological index is odd and the non-equilibrium classification for class BDI in $d = 1$ is $\mathbb{Z} \rightarrow \mathbb{Z}_2$. As such, the memory stored in the associated edge modes is not susceptible to mixing by low-frequency noise, and as expected the recovery fidelity is unaffected by the noise. On the other hand, models \eqref{eqDecoMdoelA} and \eqref{eqDecoMdoelC} lose their topology according to our classification (Table \ref{tabClassification}). In these cases, the recovery fidelity decays until the states which started with opposite qubit values become indistinguishable from one another. There is thus no way of extracting the qubit in these cases where the system topology is destroyed by non-equilibrium effects.

Simple arguments can be applied to estimate the rate at which the fidelity decays in Figure \ref{figDecoherence}. We are generally working in the regime where $E_\text{maj} \ll V,\, T^{-1} \ll E_\text{gap}$, where $E_\text{maj}$ is the energy of splitting of the Majoranas (exponentially small in the system size), $V \sim \sqrt{\langle \eta(t)^2 \rangle}$ characterizes the energy scale of the noise term, $T \sim \Gamma^{-1}$ is the time scale over which the signal fluctuates, and $E_\text{gap}$ is the energy gap of the system. In such a regime, the degenerate adiabatic theorem applies, as described in Section \ref{secLocalMix} (this is independent of the ratio of $V$ and $T^{-1}$). As such, all mixing effects are purely geometric since any dynamical phases within the low-energy subspace are on the order of $E_\text{maj}$. For small $V$, the dependence of the Berry curvature on the instantaneous value of $\vec{\eta}(t)$ can be neglected \footnote{This is valid as long as the Berry curvature at $\vec{\eta} = 0$ is non-zero, otherwise we should expand $\Omega$ to lowest non-zero order in $\vec{\eta}$. Although the fine-tuned `ideal' models such as {\protect \eqref{eqHamIdeal}} have zero Berry curvature at the fine-tuned point, generic models including Hamiltonians {\protect \eqref{eqModels}} do have a non-zero Berry curvature.}, so that the angle of mixing between two Majoranas after a time $t$ is $\theta(t) \sim \Omega A(t)$, where $A(t)$ is the signed area swept out by the $\vec{\eta}$ vector over a time $t$. Clearly, if we only have one noise source, $\Dim \vec{\eta} = 1$, then this area is identically zero -- this is why we included two (uncorrelated) noise sources in the above. With multiple noise channels, the root mean square area is roughly $\sqrt{\langle A(t)^2 \rangle} \sim V^2(t/T)$. Therefore at short times, the mean-square angle of mixing grows with time as $\sqrt{\langle \theta(t)^2 \rangle} \sim \Omega V^2 t \Gamma$. The Berry curvature will be on the order of the inverse square of the energy scale of the Hamiltonian $\sim E_\text{gap}^{-2}$. Hence the timescale over which the Majoranas mix is $\tau_\text{d} \sim E_\text{gap}^2 V^{-2} \Gamma^{-1}$. After averaging over the noise, this leads to a decoherence in the fidelity as $e^{-t/\tau_\text{d}}$.

Although the above arguments are specific to 0D edge modes, wherein a degenerate adiabatic approximation can be reliably applied, we expect that the relationship between the non-equilibrium destruction of bulk topology listed in Table \ref{tabClassification} and the instability of edge modes against temporal fluctuations should hold generally, and in particular be observable in transport signatures.

\section{Relation to Floquet-SPT Phases \label{secFloquet}}

The non-equilibrium topological classification which we have developed in the previous sections applies rather generally to systems undergoing unitary time evolution in which the time-dependence of the Hamiltonian is arbitrary (but symmetry-respecting). There are, however, topological characterizations of non-equilibrium dynamics which apply to more specific protocols; most notably in the dynamics of periodically driven systems, where Floquet-SPT order can emerge\cite{Else2016,vonKeyserlingk2016,Potter2016,Roy2017}. It is worth understanding how our results relate to those found in that context.

In order to make connection with observables, we have only considered properties which can be inferred from the wavefunction $\ket{\Psi(t_\text{f})}$ at some instant in time $t_\text{f}$. This ensures that the topology which we refer to can be detected using, e.g.~the entanglement spectrum. However, if periodicity of dynamics is enforced (either by looking at a Floquet eigenstate, or by considering the steady state of a many-body localized system), then the micromotion over a period can also be characterized. Floquet-SPT order captures the topological properties of this micromotion which cannot be inferred at a single time during the cycle; for that reason, it is fundamentally different from our characterization. Indeed given that generic unitary evolution does not generate periodic evolution of $\ket{\Psi(t)}$, there is no discrete time-translational symmetry in our protocol which is central to the stabilization of Floquet-SPT phases.

This distinction can be seen most clearly in the dynamics of the entanglement spectrum in 1D systems. We have demonstrated that robust non-equilibrium topology of the wavefunction ensures that entanglement degeneracies are preserved at all times throughout the time evolution. In contrast, as shown in Ref.~\onlinecite{Potter2016}, the pattern of entanglement spectrum dynamics in a Floquet SPT phase is one of `charge pumping', where the entanglement energies associated with entanglement eigenstates of opposite symmetry-charge are forced to cross each other at some point during the time evolution. Looking at the entanglement spectrum at a particular time in the Floquet system, one would generically observe no degeneracies.

Although the topological characterization of periodic and non-periodic systems captures different physics, both analyses demonstrate the stark difference between equilibrium and non-equilibrium systems. It is known that Floquet SPT order can emerge independently of the static properties of the Floquet Hamiltonian, which governs the stroboscopic time-evolution of one period \cite{Rudner2013,Titum2016}. In the same manner, the wavefunction topology which we discuss can be maintained independently of the static properties of the Hamiltonian governing time-evolution. Consider as a simple example a quench in which the initial state is the ground state of a TRS-broken insulator with non-zero Chern number, but the final Hamiltonian is time-reversal symmetric. (This quench where the symmetry of the Hamiltonian changes was not covered earlier, but can be easily understood in the same manner.) The TRS of the final Hamiltonian implies that its ground state cannot be a Chern-insulating phase, but regardless the wavefunction $\ket{\Psi(t)}$ will be topologically non-trivial while it evolves under the trivial $\matr{H}^\text{f}$, analogous to the cases studied in Refs.~\onlinecite{DAlessio2015,Caio2015}. Rather generally, we find that the topological properties of a time-evolving wavefunction are completely independent from the static topological indices associated with the governing Hamiltonian.

%We can also obtain an estimate for the long-time value of the fidelity in the model of Eq.~\eqref{eqDecoMdoelB}. In this case, the Berry curvature is identically zero, however there is a small amount of decoherence due to the imperfect overlap of the initial Majorana wavefunction with the instantaneous Majorana mode; i.e.~we need to account for the fact that we are not following a perfect loop in $\vec{\eta}(t)$-space. This leads to a small mixing angle $\theta(t) \sim \int_{\eta(0)}^{\eta(t)} \diff \vec{\eta} \cdot \vec{\mathcal{A}}(\vec{\eta})$. This angle does not accumulate as $t \rightarrow \infty$, but fluctuates around zero with amplitude on the order of $V \sqrt{\langle\mathcal{A}^2\rangle}$. The typical value of the Berry connection is on the order of $E_\text{gap}^{-1}$. Therefore the noise-averaged long-time fidelity $= \cos \theta(t\rightarrow \infty)$ will approach $1 - \alpha (V/E_\text{gap})^2$ for some dimensionless constant $\alpha$.

\section{Conclusion \label{secConclusion}}

We have developed a formalism by which the topological properties of pure many-body wavefunctions far from equilibrium can be understood. Importantly, we show that such systems possess a non-equilibrium topological classification which can differ from that in equilibrium. This approach was applied to non-interacting fermionic systems under the `ten-fold way' in arbitrary spatial dimension, which led to our central result, Table \ref{tabClassification}. Robustness to disorder and weak interactions are naturally incorporated in our results, as is familiar from equilibrium topology. The results can be understood using two complementary perspectives: one in terms of bulk properties and one in terms of boundary theories, the latter of which can be probed using the entanglement spectrum. The physical implications of our results were discussed, and we demonstrated that our classification correctly predicts which topological zero-energy modes are unstable to external fluctuating perturbations. This naturally has consequences for the usage of such zero-modes as a topological qubit memory.

Recent results on 1D spin chains \cite{McGinley2018} also indicate that non-equilibrium classifications could be constructed for wider classes of systems that feature strong interactions and/or spatial symmetries. Additionally, understanding further implications of our classification for experimentally relevant settings remains an important challenge. In analogy to our results on decoherence of zero-energy bound states, we expect that the effect of external time-dependent perturbations on bulk and edge mode transport signatures, such as those studied in Ref.~\onlinecite{Vayrynen2018} will be directly linked with our non-equilibrium classification.

\acknowledgements{This work was supported by an EPSRC studentship and grants EP/K030094/1 and EP/P009565/1, and by an Investigator Award of the Simons Foundation.  Statement of compliance with EPSRC policy framework on research data: All data are directly available within the publication.}

%We have studied the topological properties of many-body wavefunctions resulting from quantum quenches between Hamiltonians, focusing on topological insulators and superconductors. When initial and final Hamiltonians possess the same set of symmetries, it was shown that the time-evolved state can have a reduced symmetry compared to the governing Hamiltonians. Topological states which depend on these symmetries can thus be destroyed out of equilibrium. We constructed a classification of all topologically distinct states which can arise after a quench as a function of spatial dimension and symmetries of the Hamiltonians -- this is shown in Table \ref{tabClassification}.  Our work elucidates the role of topology in coherent non-equilibrium scenarios.
%\nocite{apsrev41Control}
%\bibliographystyle{apsrev4-1-prx}
\bibliography{topology_ooe_class}

\end{document}